\documentclass[12pt,preprint]{aastex}

\shorttitle{New line--driven stellar wind solutions}
\shortauthors{M. Cur\'{e}}
\begin{document}
\title{The influence of rotation in radiation driven wind from hot  
stars: New solutions and disk formation in Be stars.} 
\author{Michel Cur\'{e}} 
\affil{Departamento de F\'{\i}sica y Meteorolog\'{\i}a \\ 
Universidad de Valpara\'{\i}so, Casilla 5030, Valpara\'{\i}so, Chile}  
\authoremail{michel.cure@uv.cl} 
\begin{abstract} 
The theory of radiation driven wind including stellar rotation is 
re-examined. After a suitable change of variables, a new equation 
for the mass loss rate is derived analytically. The solution of this equation
remains within $1\%$ confidence when compared with numerical solutions.
Also, a non-linear equation for the position of the critical (singular) point
is obtained. This equation shows the existence of an additional critical
point, besides the standard m--CAK critical point.
For a stellar rotation velocity larger than $\sim$ 0.7  -- 0.8 $V_{breakup}$, there exists only one critical point, located away from the star's surface.
Numerical solutions crossing through this new critical point, are 
attained. In these cases, the wind has a very low terminal velocity and therefore a higher density wind. Disk formation in Be stars is discussed in the frame of this new line driven stellar wind solution.
\end{abstract} 
 
\keywords{ 
hydrodynamics --- methods: analytical--- stars: early-type --- stars: mass-loss
--- stars: rotation --- stars: winds, outflows} 
 
\section{Introduction} 
The theory of radiation driven winds has been very successful 
describing the observed terminal velocities and mass loss rates from hot 
stars. After the pioneering work of Castor, Abbott and Klein (1975, hereafter 
CAK) who realized that the force due to line absorption in a rapidly expanding 
envelope can be calculated using the Sobolev approximation (Castor 1974). They 
developed a simple parameterization of the line force and were able to 
construct an analytical wind model.
 
Simultaneously and independently, Friend and Abbott (1986) and Pauldrach, Puls 
and Kudritzki (1986) (hereafter FA and PPK, respectively) calculated the 
influence of the finite cone angle correction on the dynamics of the wind.
They found a better agreement between the improved or modified theory
(hereafter m-CAK) and the observations for both $\dot{M}$ (mass loss rate) and
$v_{\infty }$ (terminal velocity) in a large domain in the Hertzsprung-Russell
diagram. Furthermore, Kudritzki et al. (1989, hereafter KPPA) developed analytical 
formulae for the localization of the critical point, mass loss rate and terminal
velocity with an agreement within 5\% for $v_{\infty }$ and 10\% for $\dot{M}$,
when compared with numerical calculations. 

The influence of the star's rotation was investigated by Castor (1979) and 
Marlborough and Zamir (1984). Both studies concluded that the effect of 
adding a centrifugal force term in the CAK equation results in a lower 
terminal velocity wind, but the mass loss rate is not substantially 
affected. Similar results were founded by FA and PPK.
 
Concerning Be Stars, the m-CAK theory gives a good description for the polar
wind, but it fails to account for a slowly accelerating flow, like the one in
the equator of Be stars. Therefore several additional mechanisms has been
proposed, based on the radiatively driven wind theory to explain the equatorial
flow of these objects. The influence of magnetic fields and rotation were
studied by Friend and McGregor (1984, see review from Cassinelli 1998). A model 
with azimuthal symmetry, rotation and viscous force was developed by de Araujo 
et al.~(1994) and Koninx \& Hearn (1992) incorporated sound waves in the wind
dynamics of Be stars. A model driven mainly by thin lines has been proposed by 
de Araujo~(1995), he obtains an outflow with a shallow expansion and a large 
mass flux when the line force parameters are considered as {\it free} and not in
a self-consistently mode (Pauldrach 1989).
A 2-D hydrodynamical model from Bjorkman and Cassinelli
(1993) and Owocki et al. (1994) concluded that a meridional current may be
responsible for the concentration of matter towards the equator, forming a wind compressed disk ("WCD").
An assumption of the WCD model, that the line--force is strictly central was released by Owocki, Cranmer \& Gayley (1996). They conclude that non--radial line--force components together with gravity darkening effects can inhibit the formation of a WCD structure.

Despite all the efforts, the most important problem that pervades all
these models, is the strong equatorial expansion that they exhibit.
Not only the terminal velocity is too high ($\sim$  1000 $km$ $sec^{-1}$) but also there is a sharply increase of the velocity field. The fitting of observed $H_{\alpha
}$ line profiles requires terminal velocities of about 200 $km$ $sec^{-1}$ or
less (Poeckert \& Marlborough 1978), while from Fe II line profiles, Hanuschik~(1994, see also Waters \& Malborough 1994) concluded that the disk expansion velocity must not be much larger than the Doppler width.

When rapid rotators are theoretically studied (such as Be stars), many authors have reported the appearance of {\it{numerical problems}} when the rotational speed is about 0.8 times the break-up speed, see e.g., FA, Poe and Friend (1986), de 
Araujo and de Freitas Pacheco (1989), Boyd \& Marlborough (1991).

In view of the previous comments our purpose is to perform a re-analysis including the rotational centrifugal force term in radiation driven wind theory.

In section \ref{sec-HYD}, an analytical treatment with the inclusion of the 
star's rotational speed is carried on. In section \ref{sec-LCP}, after
a suitable coordinate transformation, exact formulae for the location
of the critical (singular) point(s) and for mass loss rate are obtained.
The roots of the critical point function defines the number of singular
(critical) points and their location. We show the existence of a new family of singular points in addition to the standard one (m--CAK solution family).
A simple, approximative, treatment for the location of the singular
point(s) and the value of the corresponding mass loss rate (eigenvalue) 
is introduced in section \ref{sec-RAPP}. Furthermore, in this section, numerical 
and analytical results for the standard m--CAK solution are compared, showing the
high confidence of this analytical approximation. Section \ref{sec-NEW} is
devoted to find a numerical solution from the momentum equation of the wind,
that starts at the stellar and surface reaches infinity after passing through a
critical point that belongs to this new family of singular points. Also a
comparison between numerical and analytical result are performed in
order to show the accuracy of this analytic treatment. Finally the
applicability of this model to explain disk--formation in Be--Stars 
is discussed in section \ref{sec-DF}. Conclusions and future lines of research 
are summarized in section \ref{sec-CN}. 
 
\section{Hydrodynamic formulation \label{sec-HYD}} 
The standard model for radiation driven stellar winds considers one component 
isothermal fluid in a stationary regime with spherical symmetry, neglecting the
effect of viscosity, heat conduction and magnetic field (CAK, FA, PPK).
The continuity 
equation reads:  
\begin{equation} 
4\pi r^{2}\rho v=\dot{M}  \label{2.0} 
\end{equation} 
and the momentum equation is given by:  
\begin{equation} 
v\frac{dv}{dr}=-\frac{1}{\rho }\frac{{dp}}{dr}-\frac{GM(1-\Gamma )}{r^{2}}+ 
\frac{v_{\phi }^{2}(r)}{r}+g^{line}(\rho,dv/dr,n_{E})  \label{2.1} ,
\end{equation} 
here $v$ is the fluid velocity, $dv/dr$ the velocity gradient, $\rho$ the 
mass density, $\dot{M}$ is the star's mass loss rate, $p$ the fluid 
pressure, $v_{\phi }= v_{rot} R_{\ast} / r$, where $v_{rot}$ is the star's
rotational speed at the equator, $\Gamma$ is the radiative acceleration caused
by Thomson scattering in terms of gravitational acceleration and
$g^{line}(\rho,dv/dr,n_{E})$ is the acceleration due to the lines. The standard
parameterization of the line force (Abbott 1982, PPK, FA) is: 
\begin{equation} 
g^{line}=\frac{C}{r^{2}}\;CF(r,v,dv/dr)\;\left( r^{2}v\frac{dv}{dr} 
\right) ^{\alpha }\;\left( \frac{n_{E}}{W(r)}\right) ^{\delta }  \label{2.2} .
\end{equation} 
The coefficient $C$ is given by:
\begin{equation} 
C=\Gamma GMk\left( \frac{4\pi }{\sigma _{E\;}v_{th}\;\dot{M}}\right) 
^{\alpha }\; \label{2.3} ,
\end{equation} 
here $v_{th}$ is the thermal velocity of the protons, $\sigma _{E}$ is the 
Thomson scattering absorption coefficient per density and $n_{E}$ is the 
electron number density in units of $10^{-11}cm^{-3}$, 
\begin{equation} 
W(r)= \frac{1}{2} \{ 1 - \sqrt{1-(R_{\ast }/r)^2} \} \label{2.2z} 
\end{equation} is the 
dilution factor, $CF$ is the correction factor (see appendix \ref{appB}) 
and all the other quantities have their usual meaning (see, e.g., PPK).

Introducing the following change of variables  
\begin{mathletters} 
\begin{eqnarray} 
u &=&-R_{\ast }/r  \label{2.4a} ,\\ 
w &=&v/a  \label{2.4b} ,\\ 
w'&=&dw/du  ,\label{2.4c}
\end{eqnarray}
\end{mathletters}where $R_{\ast }$ is the star's radius and $a$ is the
isothermal sound speed, 
i.e., $p=a^{2}\rho$. 

The momentum equation (\ref{2.1}) with the line 
force (\ref{2.2}) becomes: 
\begin{equation} 
F(u,w,w^\prime) \equiv \left( 1-\frac{1}{w^{2}} \right)
w\frac{dw}{du}+A+\frac{2}{u}+a_{rot}^{2}u-C^{\prime } 
\;CF\;g(u)(w)^{-\delta }\left( w\frac{dw}{du}\right) ^{\alpha 
}\ = 0 
\label{2.5} 
\end{equation} 
here 
\begin{mathletters} 
\begin{equation} 
A=\frac{GM(1-\Gamma )}{a^{2}R_{\ast }}=\frac{v_{esc}^{2}}{2a^{2}}  \label{2.5d} \;, 
\end{equation} 
\begin{equation} 
C^{\prime }=C\;\left( \frac{\dot{M} D}{2\pi}\frac{10^{-11}}{aR_{\ast
}^{2}} \right)^{\delta }\;(a^{2}R_{\ast })^{(\alpha -1)}  \label{2.5b} \;,
\end{equation} 
\begin{equation} 
g(u)=\left( \frac{u^{2}}{1-\sqrt{1-u^{2}}}\right)^{\delta }  \label{2.5c} 
\end{equation} and
\begin{equation} 
a_{rot}=\frac{v_{rot}}{a}  \label{2.5e} \;,
\end{equation} 
\end{mathletters}where $v_{esc}$ is the escape velocity, and $D$ is defined by:
\begin{equation} 
D= \frac{(1+Z_{He} Y_{He})}{(1+A_{He} Y_{He})}\frac{1}{m_{H}} \label{2.5f} \;, 
\end{equation} here $Y_{He}$ is the helium abundance relative to the hydrogen,
$Z_{He}$ is the amount of free electrons provided by helium, $A_{He}$ is
the atomic mass number of helium and $m_{H}$ is the mass of the proton.

The standard method for solving this non-linear differential equation
(\ref{2.5}) together with the constant $C^{\prime }$ (the eigenvalue of the
problem) is imposing that the solution pass through the singular (critical)
point (see CAK for details), together with a constrain on the optical depth
integral:
\begin{equation} 
\int_{R_{\ast}}^{\infty} \sigma_{E} \rho(r) dr = \frac{2}{3} 
  \label{2.5a} 
\end{equation} 

Other authors prefer to use an equivalent lower boundary condition, e.g., setting the
the density at the stellar surface to a specific value,
\begin{equation} 
\rho(R_{\ast}) =  \rho_{\ast} ,
\label{2.5ab} 
\end{equation} 
de Araujo et al.~(1989) use $\rho_{\ast 
}=10^{-11}\, g\,cm^{-3}$,
Friend \& MacGregor~(1984) use $\rho _{\ast}=1.5\times
10^{-9}\, g\,cm^{-3}$.

A critical point is located where the singularity condition, is satisfied:
\begin{equation} 
\frac{\partial}{\partial w^{\prime}}F(u,w,w^{\prime} 
)=0  \label{2.6} 
\end{equation} 
At this specific point, regularity is imposed, namely:
\begin{equation} 
\frac{d}{du} F(u,w,w^{\prime})=\frac{\partial F}{\partial u}+\frac{\partial 
F}{\partial w}w^{\prime}=0  \label{2.7} 
\end{equation} 
 
\section{Location of the Critical Point(s) \label{sec-LCP}}
In this section, after a coordinate transformation we obtain a set of analytical
equations for the eigenvalue and the location of the singular point. For the
last one, we show, that a new family of singular points exist. We also prove,
that the standard m--CAK solution vanishes for stars with high rotational
velocity.

\subsection{Coordinate transformation} 
In order to solve equations (\ref{2.5}), (\ref{2.6}) and (\ref{2.7}), we 
use the following coordinate transformation:  
\begin{mathletters} 
\begin{equation} 
Y=w\;w^{\prime}  \label{2.8a} 
\end{equation} 
\begin{equation} 
Z=w/w^{\prime}  \label{2.8b} 
\end{equation} 
\end{mathletters} 
with these two new coordinates, the equations turns to:  
\begin{mathletters} 
\begin{eqnarray} 
\left(1-\frac{1}{YZ}\right) Y & +A+ 2/u + a_{rot }^{2}u & -\,C^{\prime} 
f_{1}(u,Z)\,g(u)\,Z^{-\delta /2}\,Y^{\alpha -\delta /2} = 0 \; , \label{2.9a}\\ 
\left(1-\frac{1}{YZ}\right) Y & & -\,C^{\prime} 
f_{2}(u,Z)\,g(u)\,Z^{-\delta /2}\,Y^{\alpha -\delta /2} = 0 \; , \label{2.9b}\\ 
\left(1+\frac{1}{YZ}\right) Y &-2Z/u^{2} +a_{rot}^{2}Z\; & -\,C^{\prime} 
f_{3}(u,Z)\,g(u)\,Z^{-\delta /2}\,Y^{\alpha -\delta /2} = 0 \; , \label{2.9c}
\end{eqnarray}
\end{mathletters}derivation details and definitions of $f_{1}(u,Z)$,
$f_{2}(u,Z)$ and $f_{3}(u,Z)$ are summarized in appendix~\ref{appA}.

Solving for $Y$ and $C^{\prime}$ from the set of equations (\ref{2.9a}),
(\ref{2.9b}) and (\ref{2.9c}), we obtain:  
\begin{equation} 
Y=\frac{1}{Z}+\left( \frac{f_{2}}{f_{1}-f_{2}}\right) \left( 
A+\frac{2}{u}+a_{rot}^{2}u\right)   \label{2.10} 
\end{equation} 
and 
\begin{equation} 
C^{\prime}(\dot{M})=\frac{1}{g f_{2}}\left( 1-\frac{1}{YZ}\right) \;Z^{\delta 
/2}\;Y^{1-\alpha +\delta /2}  \label{2.11} 
\end{equation} 
These equations are generalizations of the relations founded by KPPA (see
their eq.[21] and [34] for $Y$ and eq.[20] and [44] for the eigenvalue) 
including now the rotational speed of the star.

\subsection{The critical--point function $R(u,Z)$ \label{sec-RUZ}}
It is not possible to obtain the location of the critical point from this
set of equations because we have only three equations and four unknowns
($Y$,$C^{\prime}$, $Z$ and $u$). But from equations (\ref{2.9a}),
(\ref{2.9b}) and (\ref{2.9c}), we obtain a function, $R(u,Z)$, defined by:
\begin{equation} 
R(u,Z)\equiv -\frac{2}{Z} + \frac{2Z}{u^{2}} - a_{rot}^{2}Z
+ f_{123}(u,Z)\left( A+ \frac{2}{u}+a_{rot}^{2}u\right)   
\label{2.12} 
\end{equation} 
where $f_{123}(u,Z)$ is defined by
\begin{equation} 
f_{123}(u,Z) \equiv \frac{f_{2}(u,Z)-f_{3}(u,Z)}{f_{2}(u,Z)-f_{1}(u,Z)}
 \label{2.12b} 
\end{equation}
The root(s) of this function $R(u,Z)$ gives the location of the
critical (singular) point(s) $u_{crit}$.

Notice that {\it{no approximation}} whatsoever has been used in the 
derivation of the above equations.

\begin{figure}
\plotone{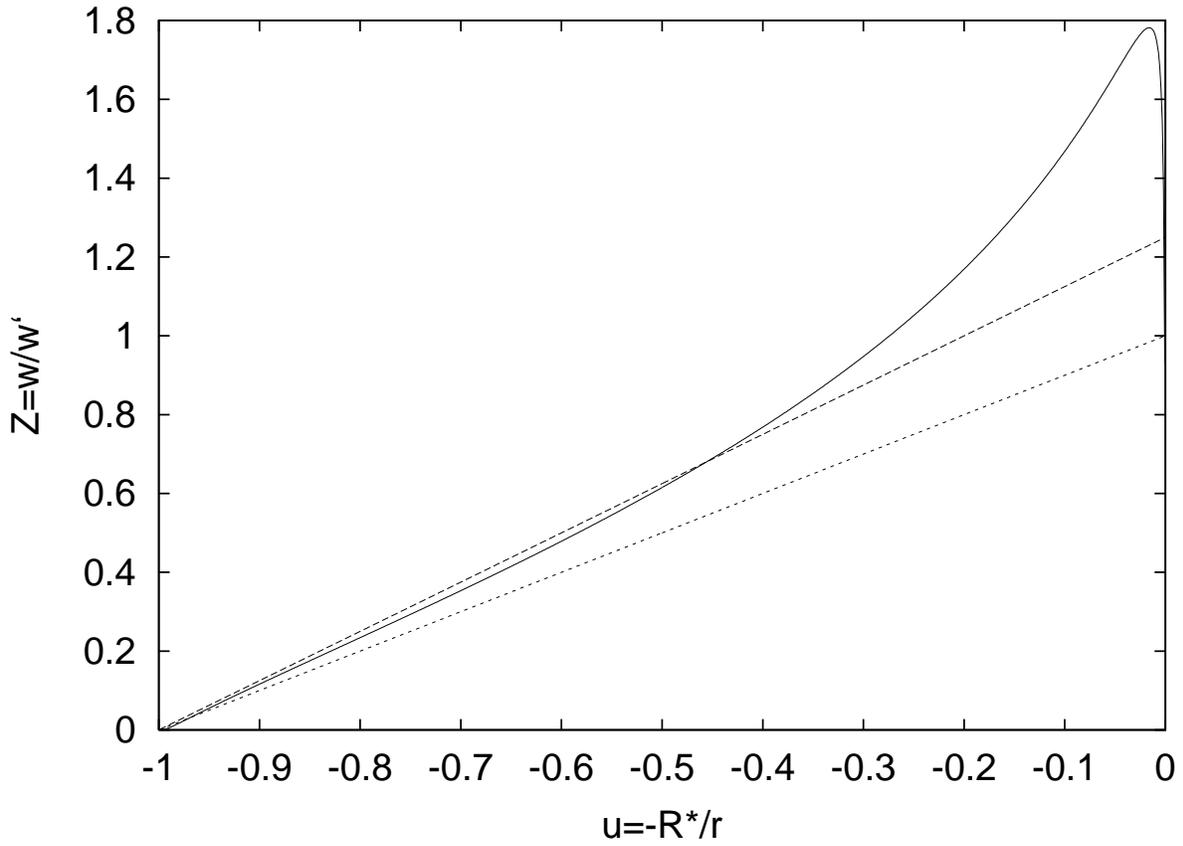}
\caption{Behavior of the variable $Z = w/w^{\prime}$ as
function of $u$ (continuous line). Two $\beta$--field approximations (straight--lines) 
for $Z=(1+u)/\beta$ are over-plotted: $\beta=0.8$ long--dashed line and 
$\beta=1.0$ short--dashed line. \label{figA}} 
\end{figure}

In order to know the range of the variable $Z$ for a typical hot--star
wind, we have performed a full numerical calculation, i.e., we solve without
any approximation: the momentum equation~(\ref{2.5}) together
with the singularity condition, equation~(\ref{2.6}), the regularity condition,
equation~(\ref{2.7}), and a lower boundary condition: equation~(\ref{2.5a}) or
equation~(\ref{2.5ab}). Our typical star is an $O5$ $V$ star with the following
stellar parameters: $T_{eff}=45000K$, $\log g=4.0$, $R/R_{\sun}=12$ and
$v_{rot}=0$, while the line force parameters are: $k=0.124$, $\alpha=0.64$ and
$\delta=0.07$. Figure \ref{figA} shows the behavior of $Z$, which ranges
between $0$ and $2$, for the whole wind. This figure also shows two
$\beta$--field approximations (see below).
\notetoeditor{Please place figure figA.eps here}

\begin{figure}
\epsscale{0.9}
\plotone{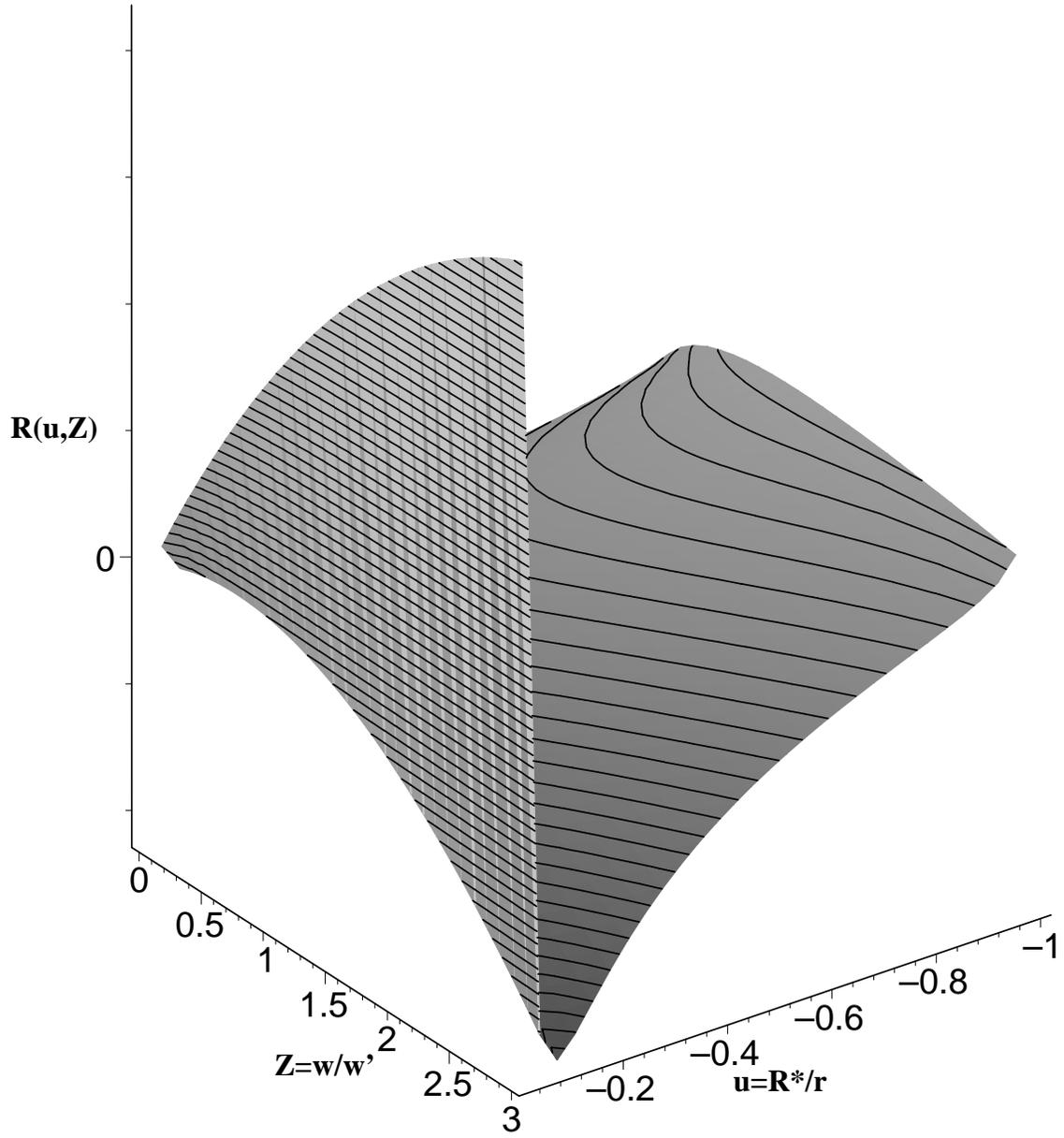}
\caption{ 
A view of $R(u,Z)$ for typical values of $u$ and $Z$ for an $O5$ $ V$ star with
$v_{rot}/v_{breakup}=0.5$. \label{fig_RUZ-a}}
\end{figure}

\begin{figure}
\epsscale{0.9}
\plotone{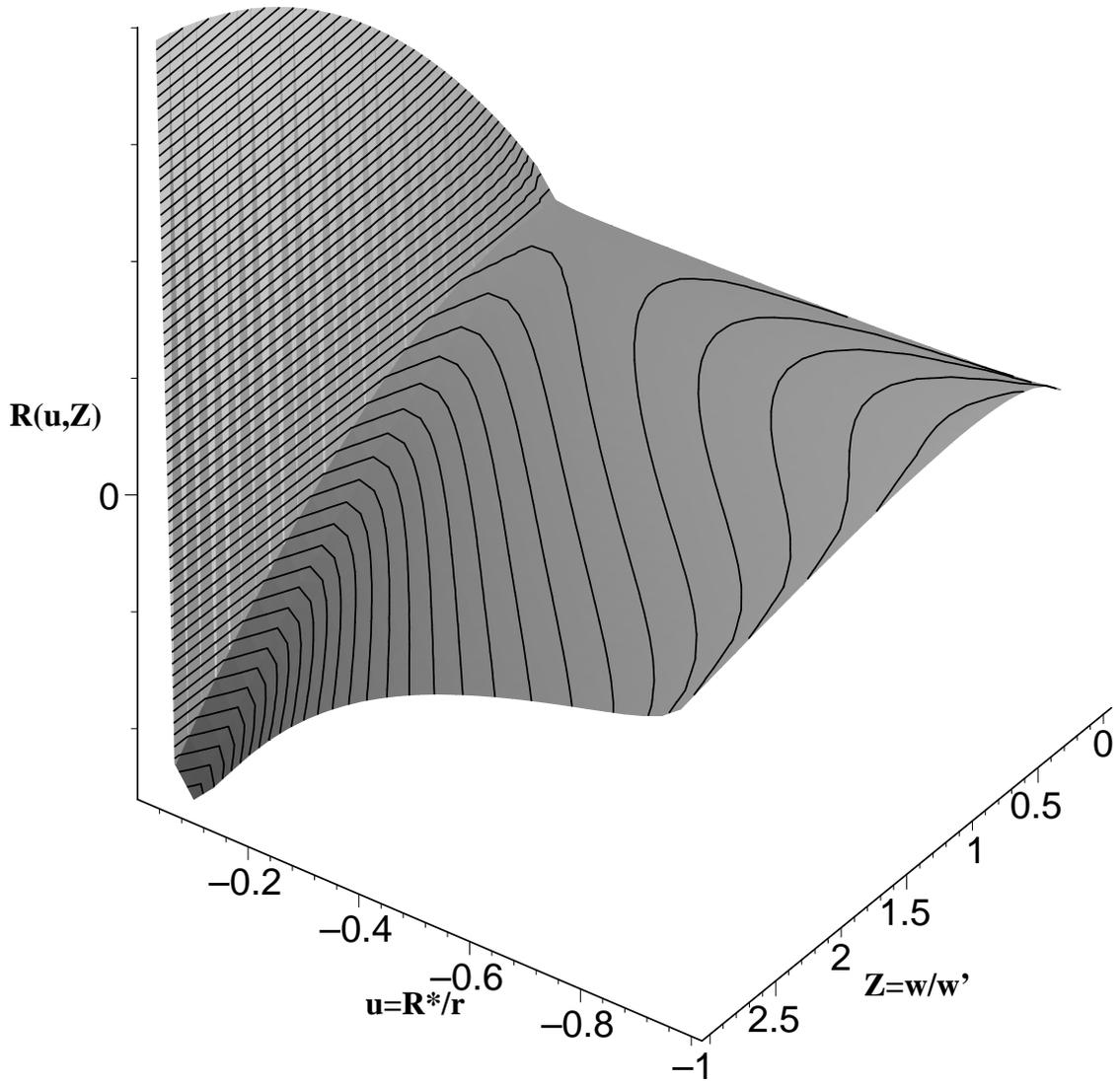}
\caption{ Same as fig.(\ref{fig_RUZ-a}) but from a different perspective.
\label{fig_RUZ-b} }
\end{figure}

\begin{figure}
\epsscale{0.9}
\plotone{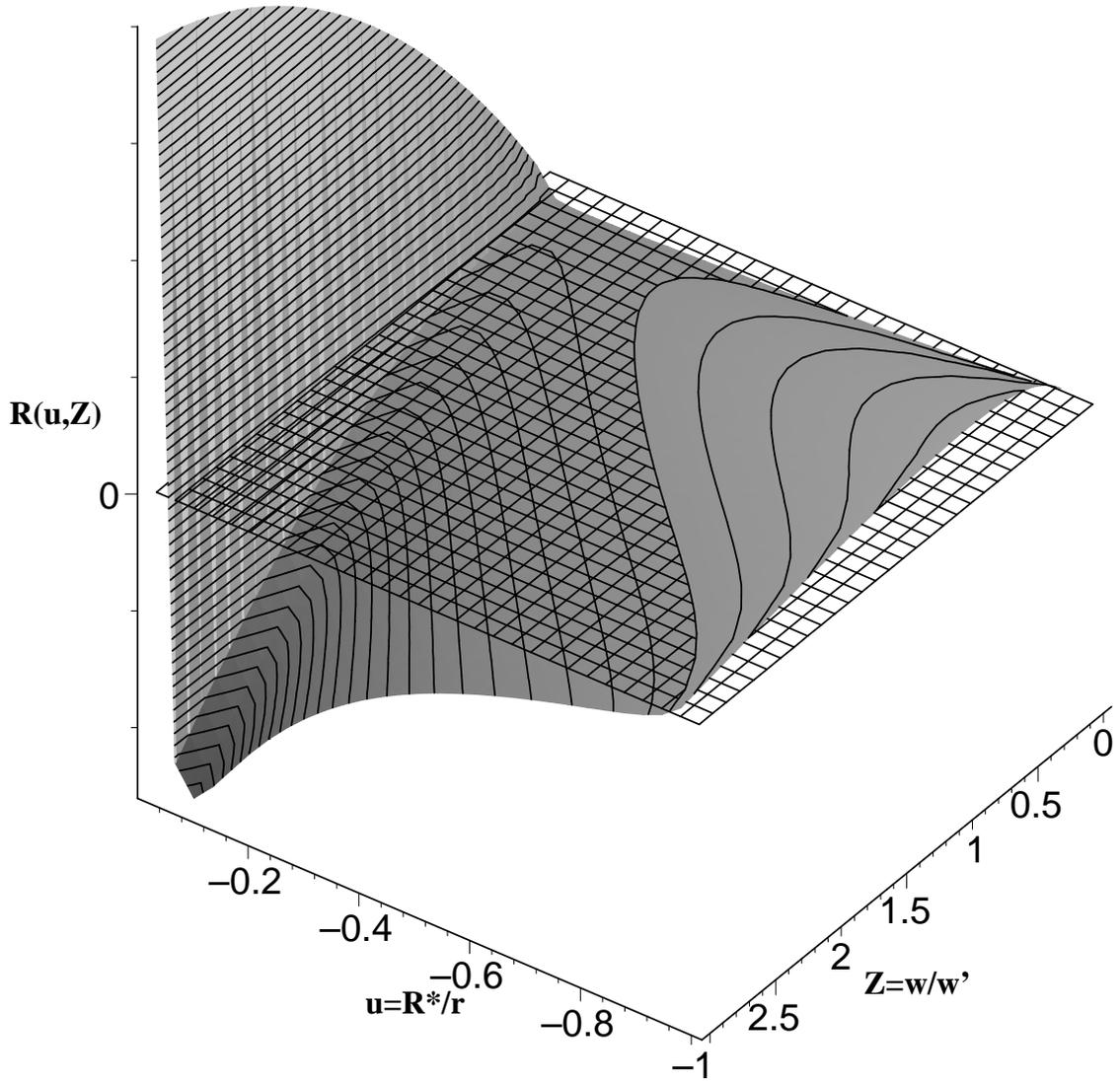}
\caption{
Same as fig.(\ref{fig_RUZ-b}), but in addition of $R(u,Z)$ the
'zero'--plane. The intersection of both surfaces gives the {\it{family}} of
solutions that satisfies simultaneously the momentum equation and both
singularity and regularity conditions. \label{fig_RUZ-c}
}
\end{figure}

Knowing now the range of the variable $Z$ we may plot $R(u,Z)$ in terms of
$u$ and $Z$. Figures \ref{fig_RUZ-a} and \ref{fig_RUZ-b} show two different 
visualization of the function $R(u,Z)$ for this $O5$ $V$ star with
$v_{rot}/v_{breakup}=0.5$.\notetoeditor{Please place figure
fig\_RUZ-a.eps and fig\_RUZ-b.eps here} 
Figure \ref{fig_RUZ-c} shows,
besides $R(u,Z)$, the {\it{zero--plane}} defined by $ R \equiv 0$.
Thus, the root(s) of $R(u,Z)$ is(are) the curve(s) defined by
the intersection of this surface $R(u,Z)$ with the {\it{zero--plane}}. Figure
\ref{fig_RUZ-c} shows both surfaces, their intersections are {\it{two}} families of critical points. The standard m--CAK family of solutions (CAK: locus of singular points) is located in the zone defined approximately by: $u \in [-1,-0.3]$ and $Z \in
[0,3]$. 

When full numerical calculations are carried on, the lower 
boundary condition, eq. (\ref{2.5a}) or eq. (\ref{2.5ab}), {\it{fixes one
point}} in this family of critical points and in this way an unique numerical
solution for the m--CAK wind is achieved. \notetoeditor{Please place figure
fig\_RUZ-c.eps and fig\_RUZ-d.eps here} 

\begin{figure}
\epsscale{0.9}
\plotone{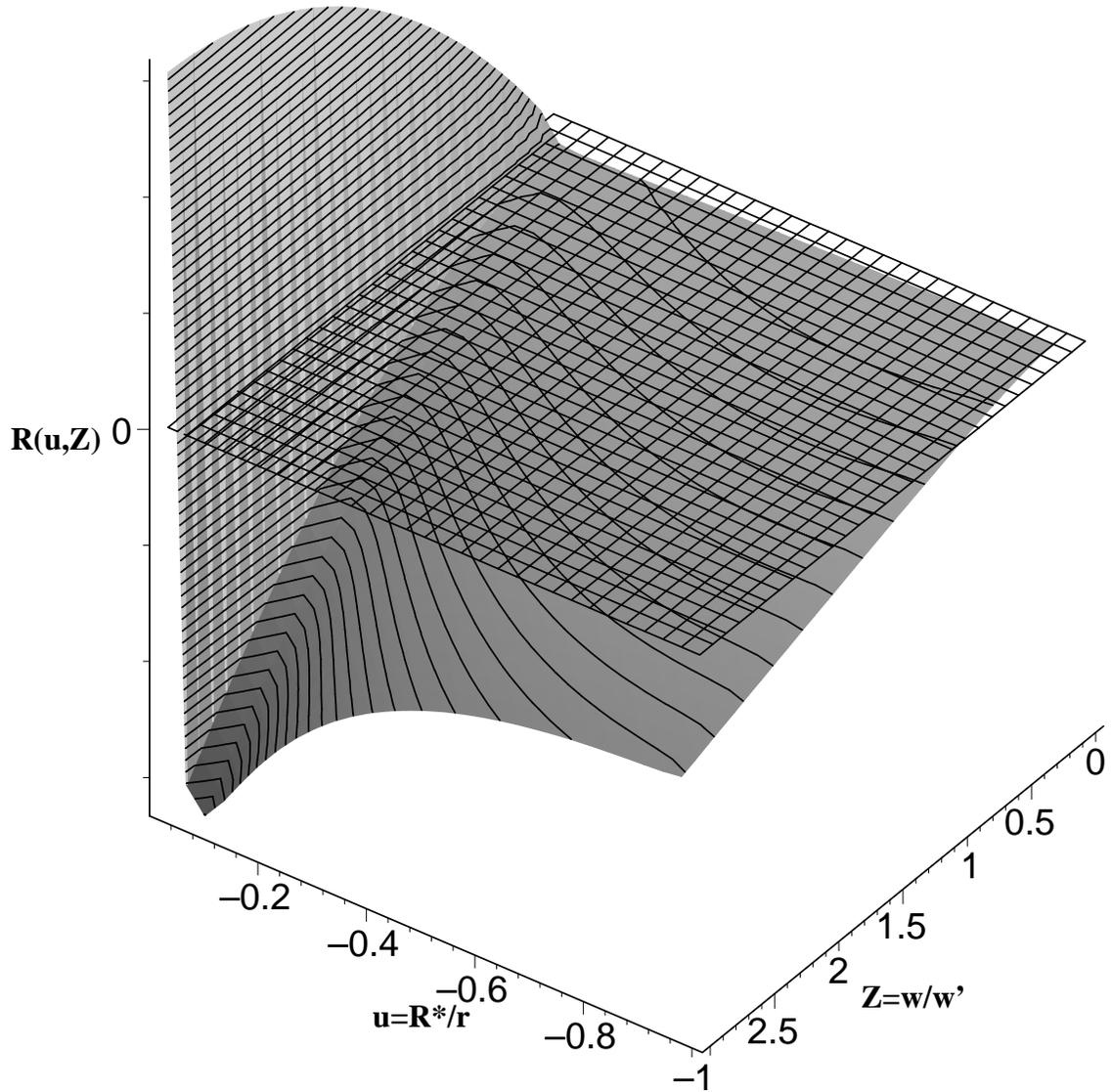}
\caption{
Same as fig.(\ref{fig_RUZ-c}), but here with $v_{rot}/v_{breakup}=0.9$. This
figure shows clearly, that no solution exists for the standard m--CAK critical
point for this rotational velocity. While a 'new' family of solutions is
founded in the region where $u \lesssim -0.2$. \label{fig_RUZ-d} }
\end{figure}

As the rotational velocity of the star increases, this m-CAK locus {\it{does not
longer intersects}} the {\it{zero--plane}} because the function $R(u,Z)$ becomes
negative in this region, as figure \ref{fig_RUZ-d} shows for the case
$v_{rot}/v_{breakup} = 0.9$. Thus, standard m--CAK solution {\it{does not
exist}} for stars with highly rotational speeds, e.g., Be--Stars. This is the reason why many authors (see .e.g., de Araujo et al. 1994, Boyd \& Marlborough 1991, FA) have
reported numerical difficulties in finding the location of the singular point
for high rotational velocities, it does not exist.

Furthermore, inspecting carefully figure \ref{fig_RUZ-d}, a {\it{new}} family
of solution is found in the region defined approximately by $u \in
[-0.2,0]$ and $Z \in [0,3]$. This family of solutions has been always present,
even for lower rotational velocities, see the case of figure \ref{fig_RUZ-c}. 

When both families of solutions are present, e.,g., for not highly rotational speeds, the family that satisfies the lower boundary condition, eq. (\ref{2.5a}), is the m--CAK locus. A study about the existence of a physical solution, that crosses through a singular point that belongs to the new family, when both families are present, is the scope of a forthcoming article. 

We are now ready to look for numerical solutions, that starts at
the stellar surface, then crosses through a singular point located at this new locus and reaches infinity, for the special case when only the new family is present, e.g., for highly rotational speeds. We will perform this task in section \ref{sec-NUM}.

Before, in the next section, a simple approximation for the function $R(u,Z)$
is introduced. This approximation allows an easy location of both, the
standard and the new singular point.

\section{Approximative Treatment \label{sec-RAPP}} 
In this section we develop an approximative treatment of the critical point
function $R(u,Z)$. After introducing a $\beta$--field approximation, the
function $R(u,Z)$ transform to $R_{app}(u)$. The zeros (roots) of this approximative function are the location of the critical points.
Knowing the location of the critical point(s) it is straightforward to calculate
the eigenvalues or equivalently the mass loss rate of the star. This
approximative calculations are in much better agreement with full numerical
calculations than previous analytical treatments and includes the influence
of star's rotation.

\subsection{The approximative critical point function $R_{app}(u)$}
From numerical calculations, PPK found that the velocity field, for stars with effective temperatures between $40.000K$ and $50.000K$ can be
approximated to the so called $\beta$--field approximation, namely:
\begin{equation} 
v=v_\infty (1+u)^{\beta}  
\label{2.12e} 
\end{equation} 
with $\beta=0.8$. This relationship is
broadly used for stellar wind diagnostics and  it is justified a posteriori by
the quality of the results achieved (Kudritzki \& Puls, 2000).

Applying this approximation for the $Z$ variable, it becomes:
\begin{equation} 
Z=\frac{(1+u)}{\beta} \label{2.12f} 
\end{equation} 
Figure \ref{figA} shows the $Z$ vs. $u$ profile from numerical calculations and  the behavior of eq.~(\ref{2.12f}) for two different values of the $\beta$ parameter ($\beta=0.8$ and
$\beta=1.0$). This approximation holds, for the $Z$ variable, up to $\sim 2.5
R_{\ast}$ ($u \sim -0.4$) for $\beta=0.8$ and up to $\sim 1.1 R_{\ast}$ ($u
\sim -0.8$) for $\beta=1.0$.

\begin{figure}
\plotone{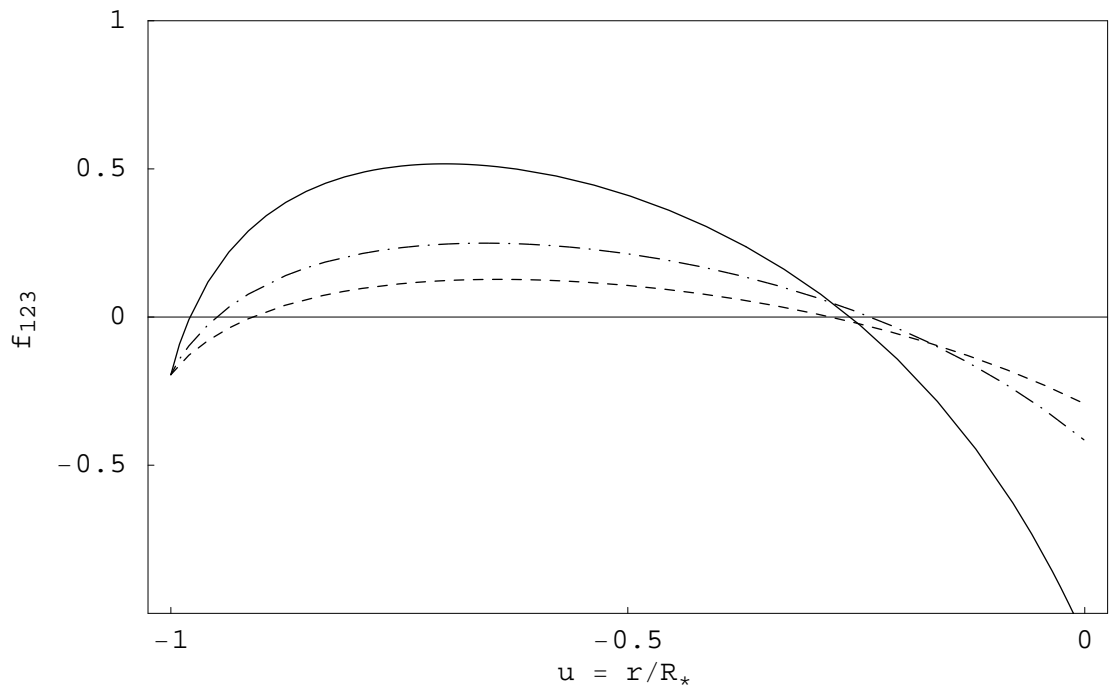}
\caption{
Function $f_{123}(u)$ vs. $u$ for different values of
$\beta$ with line--force parameters: $\alpha=0.640$ and $\delta=0.07$.
Continuous line: $\beta=1$; dashed--dotted line: $\beta=2$ and dashed
line $\beta=3$ \label{fig_f123} }
\end{figure}

Thus, replacing eq.~(\ref{2.12f}) in eq.~(\ref{2.12}), $R(u,Z)$ transform to
$R_{app}(u)$, defined as:
\begin{equation}
R_{app}(u) \equiv -\frac{2\beta}{(1+u)}+\frac{2(1+u)}{u^{2}\beta
}-a_{rot}^{2}\frac{
(1+u)}{\beta }+f_{123}(u)\left( A+\frac{2}{u}+a_{rot}^{2}u\right)
\label{2.13}
\end{equation}
The function $f_{123}(u,Z)$ transforms to $f_{123}(u)$, and its behavior
is shown in figure \ref{fig_f123} for different values of the free--parameter
$\beta$. \notetoeditor{Please place figure FIG\_f123.eps here} Once the root(s)
of $R_{app}(u)$ in the interest domain (interval $-1\leq
u\leq 0$ or $R_{\ast}\leq r\leq \infty $) is(are) obtained, the values of $Y$
and $C^{\prime}=C^{\prime}(\dot{M})$ are obtained from equations (\ref{2.10}) and (\ref{2.11}) respectively.

\subsection{An Example: CAK with Rotation \label{sec-CAK}} 
It could be thought, that this simple case has only an academic 
interest, but as we shall see later, that these results contribute to the understanding of the dynamic of rapid rotators. 
 
For the CAK model (with $\delta=0$), the  $f's$ functions are: $f_{1}=1$, $ 
f_{2}=f_{3}=\alpha$ and $f_{123}(u)=0$, then the $R_{app}(u) (\equiv
R_{CAK}(u))$
function reads:  
\begin{equation} 
R_{CAK}(u) \equiv -\frac{2\beta}{(1+u)}+\frac{2(1+u)}{u^{2}\beta
}-a_{rot}^{2}\frac{(1+u)}{\beta }  
\label{2.14} 
\end{equation} 
this is a fourth order polynomial in $u$.
Contrarily to the expected, the non--rotational case, $R_{CAK}(u)$ does not depend neither on the stellar parameters: $T_{eff}$, $M_{\ast}$, $R_{\ast}$, nor on the line--force parameters ($k$, $\alpha$, $\delta$), it explicitly depends on
$\beta$. Furthermore, for the rotational case it depends on the stellar
parameters and the free-parameter $\beta$, but not on the line-force
parameters.

CAK and KPPA have shown that the behavior of the velocity profile is given approximately by:
\begin{equation} 
v(u) = v_{\infty} \sqrt{1+u}  
\label{2.14b} 
\end{equation}

\begin{figure}
\epsscale{0.8}
\plotone{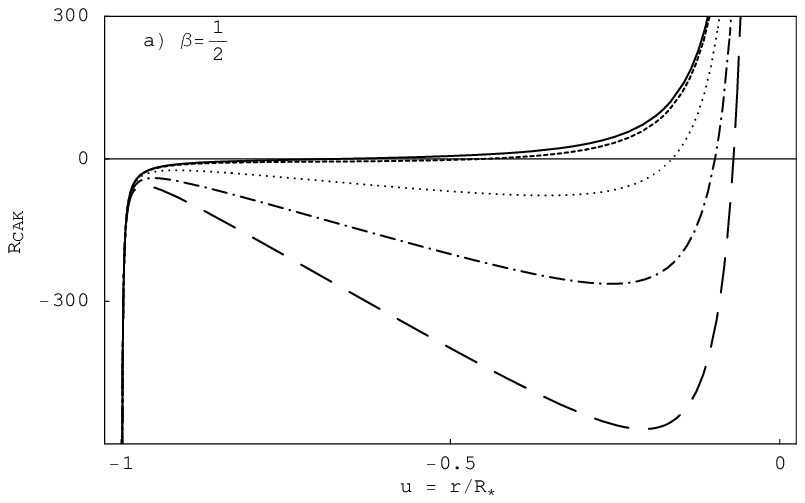}
\plotone{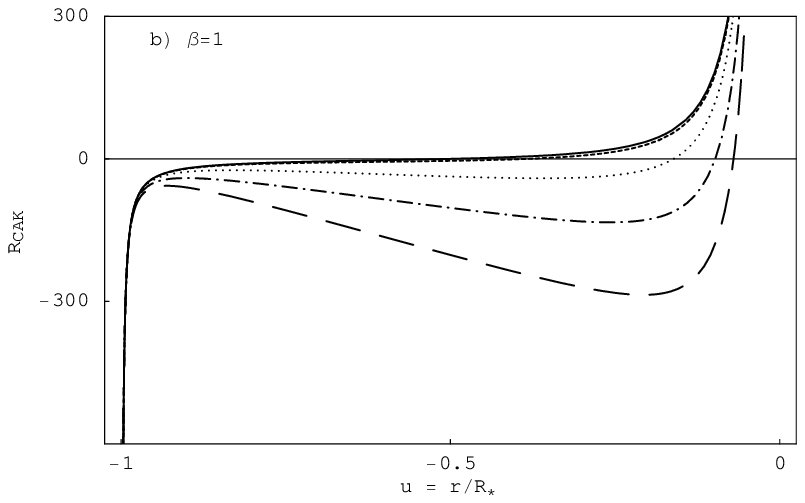}
\caption{Function $R_{CAK}(u)$ plotted against $u$ for
$v_{rot}/v_{breakup} = 0.0$, continuous line; $0.1$, dashed line; $0.3$, dotted line; $0.5$, dashed--dotted line; $0.7$, long--dashed line. Figure $a$ is for $\beta=1/2$ and figure $b$ is for $\beta=1$. The location of the singular point is when $R_{CAK}(u)=0$.\label{fig_RCAK} }
\end{figure}

This results was obtained from approximate solutions for a non--rotating CAK
wind, (see CAK eq.[47] or KPPA eq.[24]). This relationship corresponds to a
$\beta$--field with $\beta=1/2$. \notetoeditor{Please place figure
FIG\_RCAK_a.eps and FIG\_RCAK_b.eps here} Figure \ref{fig_RCAK} shows the function  $R_{CAK}(u)$, for
this $O5$ $V$ test star, for different values of $v_{rot}/v_{breakup}$ and
$\beta$.
From this figure, it is clear that {\it only one} critical point exists in the
domain of interest. Bjorkman (1995) arrived to the same conclusion in
his topological study of the non-rotating CAK model.

As the rotational speed increases, figure \ref{fig_RCAK} shows that the location of the CAK singular point is shifted downstream, this is the same conclusion reached by Castor~(1979) and Marlborough \& Zamir~(1984).

The zero ($u_{crit}$) of equation~(\ref{2.14}) in the interval $-1\leq 
u\leq 0$ is giving by the analytical formula:
\begin{equation} 
u_{crit}={\frac{1}{2}}(t_{7}-t_{8}-1)  
\label{2.15} 
\end{equation} 
where the {\it t--coefficients} are: 
\begin{mathletters} 
\begin{eqnarray} 
t_{0} &=&\left( -2+{a_{rot}^{2}}+2\,{\beta ^{2}}\right) \\ 
t_{1} &=&432\,{a_{rot}^{2}}-216\,{a_{rot}^{4}}+216\,{a_{rot}^{2}} 
\,t_{0}+2\,{{t_{0}}^{3}} \\ 
t_{2} &=&{{\left(t_{1}+\sqrt{-4 t_{0}^{6}+t_{1}^{2}}\right)}^{{\frac{1}{3}}}}\\
t_{3} &=&{\frac{{2^{{\frac{1}{3}}}}\,{{ 
t_{0}}^{2}}}{3\,{a_{rot}^{2}}\,t_{2}}} \\ 
t_{4} &=&{\frac{t_{2}}{3\,a_{rot}^{2}\,2^\frac{1}{3}}} \\ 
t_{5} &=&-8+{\frac{32}{{a_{rot}^{2}}}}+{\frac{8\,t_{0}}{{a_{rot}^{2}}}} 
\\ 
t_{6} &=&\frac{2\,t_{0}}{3\,{a_{rot}^{2}}} \\ 
t_{7} &=&\sqrt{1-t_{6}+t_{3}+t_{4}} \\ 
t_{8} &=&\sqrt{2-2t_{6}-t_{3}-t_{4}+{\frac{t_{5}}{4\,t_{7}}}} 
\label{2.15b} 
\end{eqnarray} 
\end{mathletters} 
For the simple case of ${a_{rot}=0}$, a straightforward and {\it 
exact solution} for equation (\ref{2.12}) is 
\begin{equation} 
u_{c}=-Z_{c} 
\label{2.15c} 
\end{equation} 
and after applying the approximation $Z=(1+u)/\beta$ with $\beta =1/2$, the
critical point is founded at $u_{c} = - 2 / 3 $ or $r_{c}=1.5$
$R_{\ast}$, as CAK and KPPA previously obtained using different approaches. 

Table~\ref{tab-CAK} shows the accuracy of this analytical
approximation. \notetoeditor{Please place table 1 here} 
Columns 2--4 and 6--8 show, for different values of $\beta$ parameter,
the location of the singular point (in terms of $R_{\ast}$), and mass loss
rates (in units of $10^{-6}$ $M_{\sun}$ $year^{-1}$) respectively. For
comparison, columns 5 and 10 give the values of $r_{crit}$ and $\dot{M}$ from
full numerical calculations\footnote{Here, full numerical calculation
solves equations~(\ref{2.5}), (\ref{2.6}), ~(\ref{2.7}) and the lower boundary condition equation~(\ref{2.5a}).}.
For the non--rotational case, the best values is $\beta=1/2$ confirming the
results of previous analyses. KPPA cooking recipe for $\dot{M}_{CAK}$ 
(see KPPA eq.~[27]) gives $\dot{M}_{CAK}= 3.074$ $10^{-6} M_{\sun}$
$year^{-1}$, same as the non--rotational case with $\beta= 1/2$.
While the value of the mass loss rate is almost independent of the value of
$\beta$, the location of the critical point is very sensitive to the value of $\beta$, but in the presence of rotation, this sensitivity disappears.
The reason is that here the rotational term (third term in equation
~\ref{2.14}) is the dominant one, and therefore the dependence on $\beta$
becomes minimal. Furthermore, the larger the rotational speed, the
larger is the $\beta$ value that best fits the numerical calculations.
 
The terminal velocity of the wind, $v_{\infty}$, can be now
computed from direct integration of equation (\ref{2.10}). This follows because
$Y$ is the unique solution of (\ref{2.5}) or (\ref{2.9a}).

\subsection{The m-CAK Standard Solution \label{sec-mCAK}} 
Now we analyse the influence of the finite cone-angle effect. The main
difference between this approach and the one of KPPA is that they
introduced the $\beta$--field approximation {\it before} calculating the
derivatives of the singularity and the regularity conditions. Therefore they
could not obtain the equations~(\ref{2.9a}), (\ref{2.9b}) and (\ref{2.9c}). In
fact, the last two of them become the same in the KPPA approximation.
  
\subsubsection{OB stars}
\begin{figure}
\plotone{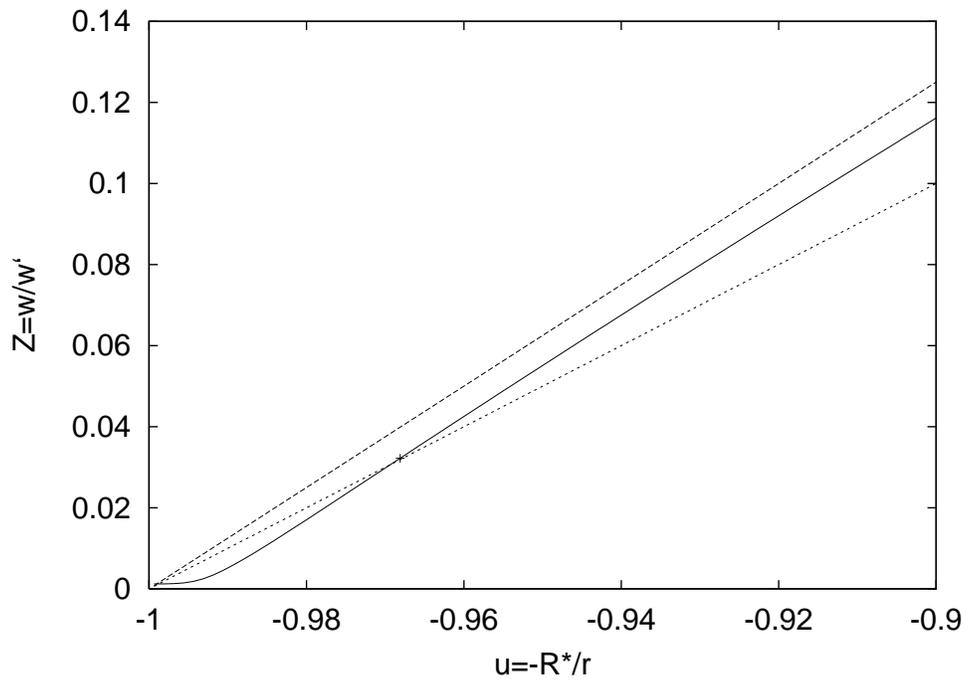}
\caption{Behavior of $Z = w/w'$ as function of $u$ (continuous--line)
in the interval $[-1.0,-0.9]$. The '+' symbol show the location of the critical
point from full numerical calculations. Two $\beta$--field approximations for
$Z=(1+u)/\beta$ are over-plotted: $\beta=0.8$ dashed--line and $\beta=1.0$
dotted--line. Clearly $\beta=1.0$ intersects $Z(u)$ almost at the
critical point. \label{figB}} 
\end{figure}

In order to obtain the critical point and the mass loss rate it
is necessary to know the value of $\beta$. As discussed above, PPK found that
a value of $\beta=0.8$ is appropriate for $OB$--Stars. In figure~\ref{figB},
same as fig.~\ref{figA} but here in the region near the photosphere, the '+'
symbol indicates the location of the critical point at the $Z(u)$ profile.
The $\beta$--field approximation with $\beta=1$ intersects almost exactly this
location showing that for this type of stars $\beta=1$ is a fairly good
approximation to calculate the critical point location and the mass loss
rate.\notetoeditor{please insert figure figB.eps and table 2}

Table \ref{tab-OB} shows a comparison between analytical and full numerical
calculations for the $O5$ $V$ test star (see stellar and line--force parameters
in section \ref{sec-RUZ}). The best agreement for both analytical,
$r_{crit}$ and $\dot{M}$, with the numerical calculation occurs for $\beta =
1$. Furthermore, this predicted analytical values agree much better with the
numerical calculations than the values given by the 'cooking recipe' from KPPA.
This recipe gives for the non--rotational case: $r_{crit} = 1.038$
$R_{\ast}$ and $\dot{M} = 1.976$ $10^{-6}$ $M_{\sun}$ $year^{-1}$ (with
$\beta=1$).
When the analytic approximation of KPPA is compared with numerical
calculations, it is found that the values of $\dot{M}$ are always
smaller than the numerical values. For the non--rotational case, our
approximation match almost exactly the  values of the numerical
calculations, but for rotational cases it still shows values slightly
less than the numerical ones.
 
\subsubsection{Cental Stars of Planetary Nebulae \label{sec-CSPN}}
Radiative driven stellar wind are also present in Cental Stars of Planetary
Nebulae (CSPN), i.e., for objects where the effect of photospheric extension is
important. Pauldrach et al.~(1988) have computed radiative driven wind models
along evolutionary tracks for these objects. They found that the the
predicted terminal wind velocities are in agreement with the observational
data, while the mass loss rate are in qualitative agreement.

In their non--LTE analysis of CSPNs, Kudritzki et al.~(1997) used a
$\beta$--field approximation for the velocity, finding that $\beta=1.5$
is the best value for these stars (see table 1 from Kudritzki et al.~1997).
Table \ref{tab-CSPN} shows a comparison between analytical and numerical
calculations for one of the models of Pauldrach et al.~(1988). The stellar
parameters for this model are: $T_{eff}=80000K$, $\log g=5.24$,
$M/M_{\sun}=0.565$ and $R/R_{\sun}=0.3$, line--force parameters are: $k=0.053$,
$\alpha =0.709$ and $\delta =0.052$. This analysis confirms that a value of
$\beta=1.5$ is a fairly good one for non--rotational CSPN, but as the
rotational velocity increases a slightly larger value must be assumed.

In this section, we have developed an analytical treatment for the location of
the critical point (standard m--CAK) and mass loss rate that improves (includes
rotation) previous investigations, and has about 1\% confidence
when is compared with full numerical calculations.

\section{Analytical Treatment of the New Wind Solution. \label{sec-NEW}} 
In previous sections, the existence of a new family of solutions,
that are different of the standard m-CAK, has been established from the
properties of the $R(u,Z)$ function. In this section we study the approximate
function $R_{app}(u)$ to get information about the location of the
singular point and the mass loss rate. 

\subsection{Critical Points \label{sec-NEWCP}} 
\subsubsection{The Standard m--CAK Critical Point \label{sec-MCAKCP}}
\begin{figure}
\epsscale{0.89}
\plotone{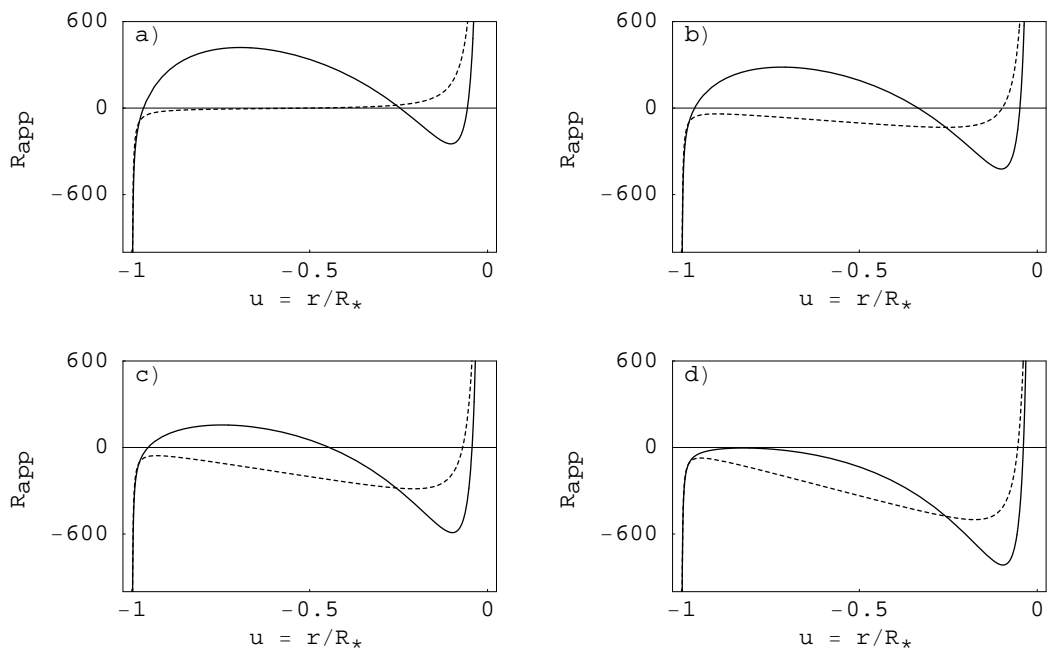}
\caption{The function $R_{app}(u)$ vs. $u$ for CAK (dashed--line) and m-CAK (continuous-line) models with $\beta=1$ for different values of the rotational speed: a) $v_{rot}/v_{breakup} = 0.0$; b)
$0.5$; c) $0.7$; d) $0.9$ for our $O5$ $V$ test star. The roots of $R_{app}(u)$ gives the approximate location of the critical points.\label{fig_RAPP_RCAK_O5V}} 
\end{figure}

Figure \ref{fig_RAPP_RCAK_O5V} shows $R_{app}(u)$, with $\beta=1$, for
an $O5$ $V$ star for different values of the star's rotational speed for both
models, CAK and m-CAK. \notetoeditor{Please place figure
FIG\_RAPP\_RCAK\_O5V.eps here} 
From this figure it is clear that the inclusion of the finite disk correction
factor in the wind momentum equation (\ref{2.5}), 'twist' the $R_{CAK}(u)$
function to $R_{app}(u)$. It is evident from this result that while
$R_{CAK}(u)$ displays only one root in the interest domain, the number of roots
of $R_{app}(u)$ depends on the rotational velocity.

For non--rotational cases (figure~\ref{fig_RAPP_RCAK_O5V}a) the CAK critical
point shifts to a location close to the stellar surface, this is the standard
m--CAK critical point. As the rotational speed increases, the location of the
m-CAK singular point is shifted downstream (FA, PPK), but differently as in the 
CAK case, it occurs up to certain rotational speed.

In order to have a better picture of the approximation $Z=(1+u)/\beta$ we have 
to recall the function $R(u,Z)$ (see, e.g., figure \ref{fig_RUZ-c}). This is a 
surface in the {\it{phase}}--space defined by the independent coordinates $u$ 
and $Z$. The approximation $Z=(1+u)/\beta$ is a vertical--plane in this 
{\it{phase}}--space that {\it{cuts}} the $R$--surface defined by equation~(\ref{2.12}). 
The projection of this intersection--curve over the vertical plane $u-R$ is 
the $R_{app}$ function showed in figure~\ref{fig_RAPP_RCAK_O5V}.

\begin{figure}
\epsscale{0.7}
\plotone{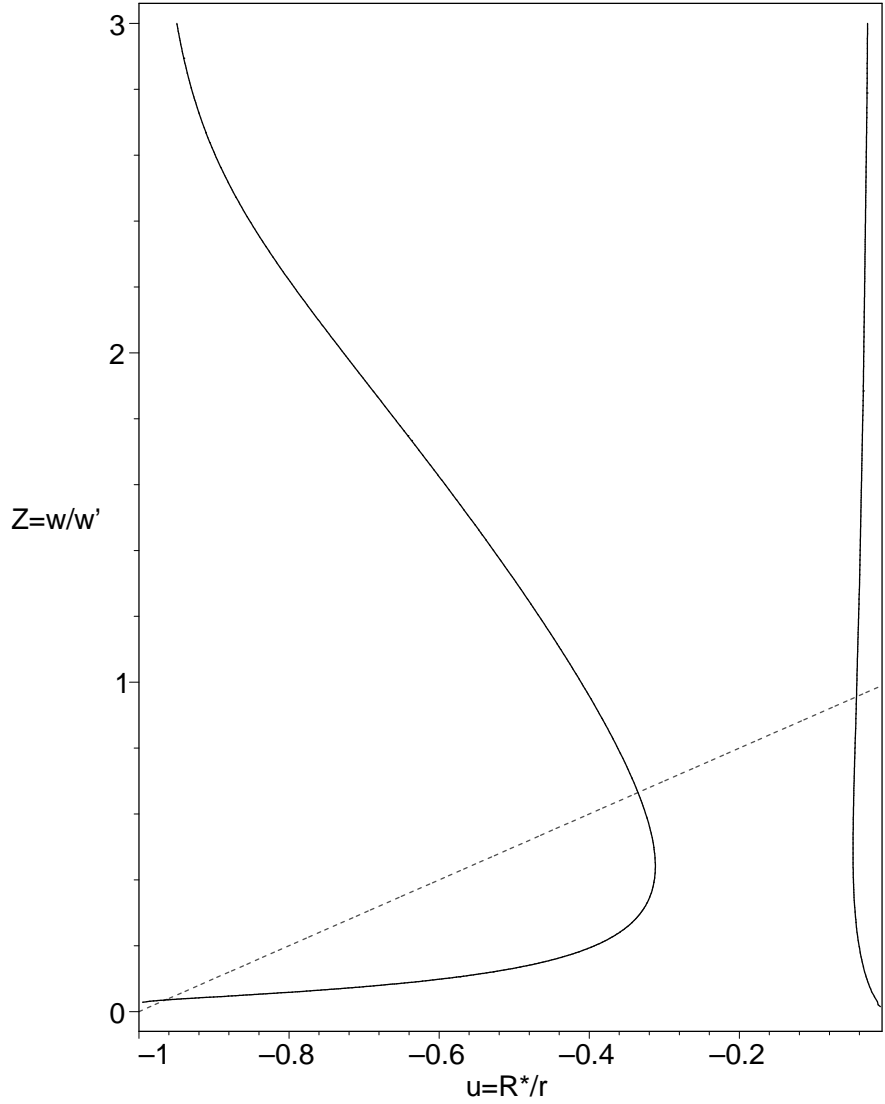}
\caption{In the {\it{phase}}-space $u-Z$, this figure shows a
countourplot of $R(u,Z)=0$ (continuous--line) and a $\beta$--field (dashed line, $\beta=1$). The continuous--lines correspond to both critical point families, m-CAK and the new one. The parameters of $R(u,Z)$ are the same as in figure (\ref{fig_RUZ-c}). \label{fig_Rcontour} }
\end{figure}

\notetoeditor{Please place figure fig\_R-contour.eps here}
Figure \ref{fig_Rcontour} is a contour plot of the $R$--surface at the {\it{zero}}--plane, 
the $u-Z$ plane. Both continuous--lines correspond to the two families (loci) of singular 
points, discussed in section (\ref{sec-NEWCP}), while the dashed--line is the $\beta$--field 
approximation ($\beta=1$). This approximation cuts the standard m--CAK family in two 
points. The standard m--CAK solution passes through the first intersection point (nearest to 
stellar surface, $u=-1$) and not through the second intersection point, as a result of imposing 
the lower boundary condition equation~(\ref{2.5a}). The second intersection point is only a 
consequence of the $\beta$--field approximation and a solution (if exists) that passes {\it{only}} 
through this point probably have no physical meaning for radiation driven winds.

\subsubsection{The {\it{New}} m--CAK Critical Point \label{subsec-NEWCP}}
In addition to the standard m--CAK critical point, figure (\ref{fig_RAPP_RCAK_O5V}) and figure~ \ref{fig_Rcontour} shows a {\it second critical point} (root of $R_{app}$), the last 
intersection--point. As the rotational speed increases, for this $O5$ $V$ star, the standard 
m-CAK critical point now {\it disappears} of the integration domain when
$v_{rot}/v_{breakup} \gtrsim 0.9$ (fig. \ref{fig_RAPP_RCAK_O5V}d, see
also figure \ref{fig_RUZ-d}). In this case, the CAK critical point is shifted far downstream 
in the wind. At the critical point, $u_{crit}$, the function $f_{123}(u_{crit}) \rightarrow 0$, 
so $R_{app}(u)$ is almost the same for CAK and m-CAK models (see equations \ref{2.13} 
and \ref{2.14}). Thus, for the fast--rotational case, equation (\ref{2.15}) gives a very good
approximation for the location of the critical point and therefore the mass loss rate, too. 

We conclude that the CAK model with rotation is applicable for high--rotational velocities 
to calculate the mass loss rate and the location of the singular point of the m--CAK model.

\begin{figure}
\epsscale{.8}
\plotone{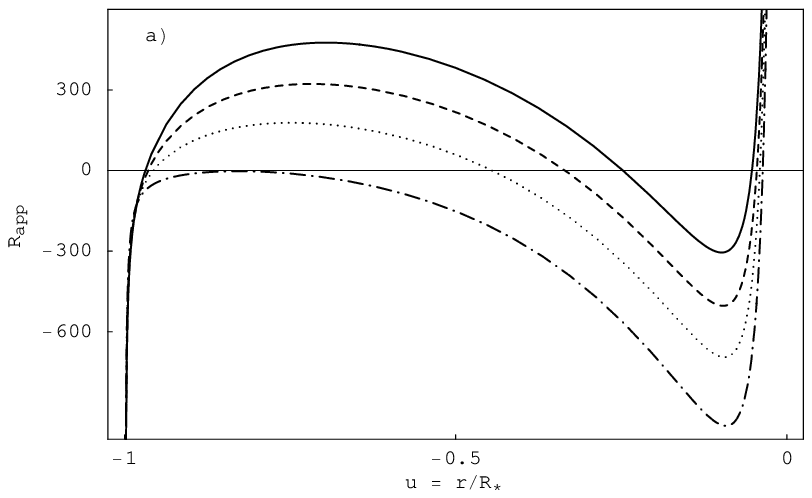}
\plotone{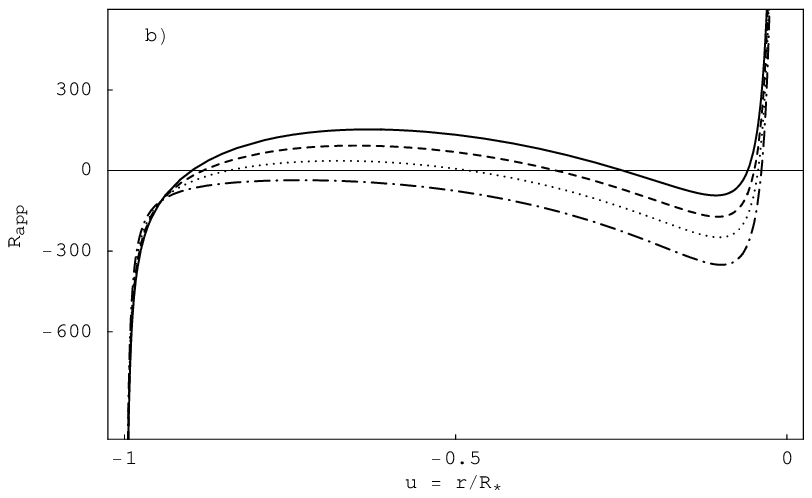}
\caption{The function $R_{app}(u)$ versus $u$ for a $B1$ $V$ star for two different
values of the free-parameter $\beta$; a) $\beta=1$; b) $\beta=2.5$.
The different rotational speeds $v_{rot}/v_{breakup}$ are: $0.0$, continuous line; $0.5$, dashed line; $0.7$, dotted line; $0.9$, dashed--dotted line. \label{fig_RAPP_Be}} 
\end{figure}

Because we are now interested in stars with high rotational velocities,
from now on our {\it{test star}} will be a typical $B1$ $V$ Star with following
stellar parameters: $T_{eff}=25000K$, $\log g=4.03$, $R/R_{\sun}=5.3$
(Sletebak et al. 1980) and line force parameters: $k=0.3$, $\alpha=0.5$, 
$\delta=0.07$. 
\notetoeditor{Please place figure FIG_RAPP_Be_a.eps and FIG_RAPP_Be_b.eps here}
For this star, figure \ref{fig_RAPP_Be} shows $R_{app}(u)$ for two different
values of $\beta$. In both cases, as the rotational speed increases,
$R_{app}(u)$ goes to a scenery where there is only one critical point, the new one.
For $\beta=2.5$, a lower value of the rotational speed is required to get to
this situation, i.e., $v_{rot}/v_{breakup} \gtrsim 0.6$, while
for $\beta=1$, the required is $v_{rot}/v_{breakup} \gtrsim 0.8$. This last
value is lower than the required for the $O5$ $V$ Star (see fig.
\ref{fig_RAPP_RCAK_O5V}).

In order to know all the families of solutions, a topological analysis is
needed, similar to the analysis performed for the non--rotating CAK
model by Bjorkman~(1995). As a first step in order to learn about this new solution, 
we carry on numerical calculations only for the fast--rotational case, i.e.,
the case where only {\it{one}} root of the $R_{app}$ function exists, the new critical point. 
It is not clear yet, that a wind solution should cross {\it only} through {\it one} critical point or there are solutions that have to cross through more than one critical points as, e.g., in the magnetic wind model of Friend \& MacGregor
(1984).

\section{Numerical Results \label{sec-NUM}}
It has been established in the previous sections that in a fast--rotational
case, the standard m--CAK critical point vanishes and there exist a second critical point. 
In this section we will obtain full numerical solutions, that starting at the star's 
surface reaches infinity passing through this new critical point.

\subsection{The Numerical Method} 
In order to solve the non--linear wind momentum differential equation 
(\ref{2.5}) for a fast--rotational case, we will not use the 'standard'
numerical method, i.e., once the singular point location, its velocity and
velocity gradient are approximatively obtained, direct integration up and
downstream are performed (e.g., using Runge-Kutta or Burlish-Stoer methods).
After this process has been done, the lower boundary condition is calculated
and the whole procedure starts again until convergence is achieved.

Instead, we use here a finite-difference method, modified for handling singular points (Nobili \& Turolla 1989). This method uses a trial solution as its initial guess. This initial solution then relaxes, using the Newton method, towards the numerical solution (see Nobili \& Turolla 1989 for details). The main advantage of this method is the fact that it is not necessary to give {\it{a priori}} the location of the critical point, as the standard method requires, so no guess value for $\beta$ is needed for solve the momentum equation.

\subsection{The Fast Solution \label{sec-FS}}
First we calculate the non--rotational case, $v_{rot }=0$, for our test star 
using $\tau = 2/3$ (eq. \ref{2.5a}) as lower boundary condition.
\notetoeditor{Please place figure fig\_vels\_be.eps here (in this
subsection)} In figure \ref{fig_vels_be} the velocity profile for this cases
(dashed--line) is shown. Numerical results are: critical point location at
$r_{crit}=1.014\,R_{\ast}$; mass loss rate $\dot{M}
= 3.178\,10^{-9}\,M_{\sun}\,year^{-1}$; and terminal velocity $v_{\infty
}=2025\,km\,sec^{-1}$.

\subsection{The Slow Solution \label{sec-SS}}
\begin{figure}
\plotone{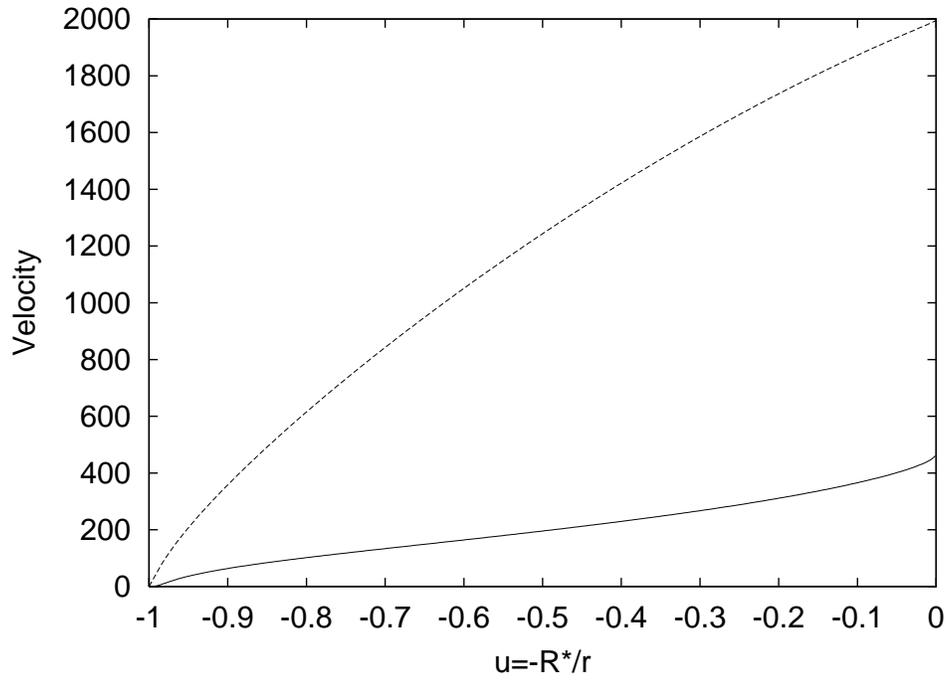}
\caption{Velocity profiles $v$ versus $u$ for a $B1$ $V$ star. Non--rotational
case: dashed line; high rotational case with $v_{rot}/v_{breakup}=0.8$:
continuous line. \label{fig_vels_be}} 
\end{figure}

We have solved numerically the fast--rotational case for $v_{rot}/v_{breakup} =
0.8$.  This new solution is  denser than the fast solution (m--CAK) and we
call it hereafter {\it{slow}} solution due to the lower value of the terminal
velocity achieved, figure \ref{fig_vels_be} also shows this profile (continuous--line).
The location of the critical point is at $r_{crit}=23.63\;R_{\ast}$,
confirming the approximative calculations of the preceding sections, that the
critical point is located far away in the wind. The terminal velocity of this
new solution is $v_{\infty }=443\,km\,sec^{-1}$. This value is a factor 4 lower
than the standard m--CAK solution (the fast--solution, dashed--line in figure
\ref{fig_vels_be}). The mass loss rate is $\dot{M} = 6.854\,10^{-9} M_{\sun
}\,year^{-1}$, approximately twice the value of the non--rotational case.

\subsubsection{The Approximation $Z=(1+u)/\beta$ \label{sub-ZB}}
\begin{figure}
\plotone{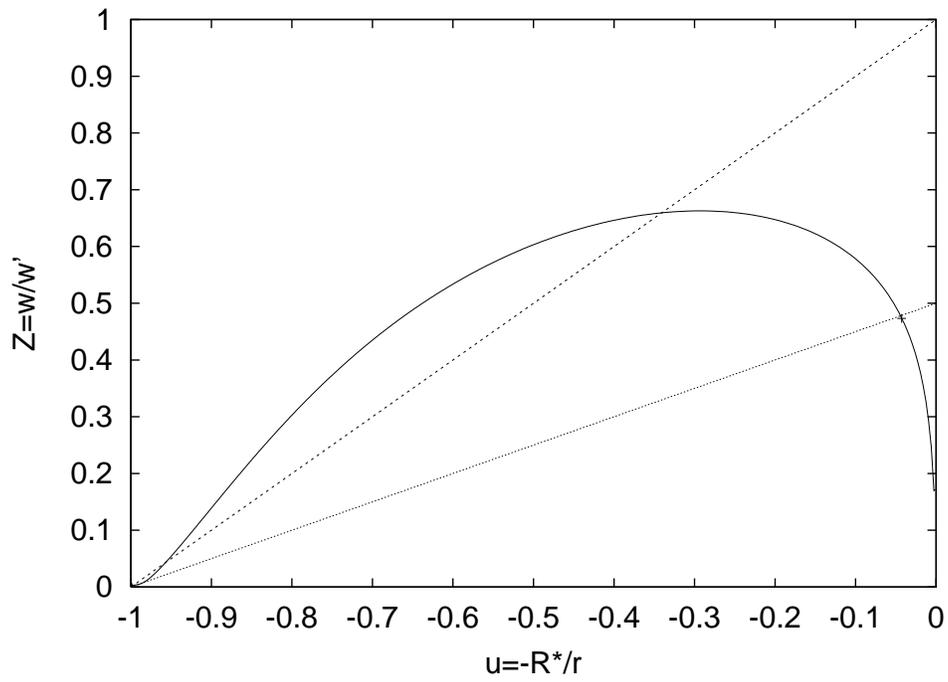}
\caption{$Z$ versus $u$ from full numerical calculations, continuous line;
The approximation$Z=(1+u)/\beta$ for $\beta=1$ dashed--line, and for $\beta=2$
dotted--line
short--dashed line. \label{fig_Z_be}} 
\end{figure}

Figure \ref{fig_Z_be} shows $Z$ versus $u$ from full numerical calculations
(continuous line). In addition, the location of the critical point is drawn with
a '+' symbol in this figure. The approximation $Z=(1+u)/\beta$ (a straight line 
in this figure) is also plotted for two different values of the
parameter $\beta$. The line with $\beta=2$ intersects $Z(u)$ close to the
location of the critical point, showing that for fast--rotational cases a value
of $\beta > 1$ matches better the location of the critical point and the mass
loss rate (eigenvalue).
\notetoeditor{Please place figure fig\_Z\_be.eps here}

\subsubsection{Functions $R(u,Z)$ and $R_{app}(u)$ \label{sub-RR}}
\begin{figure}
\plotone{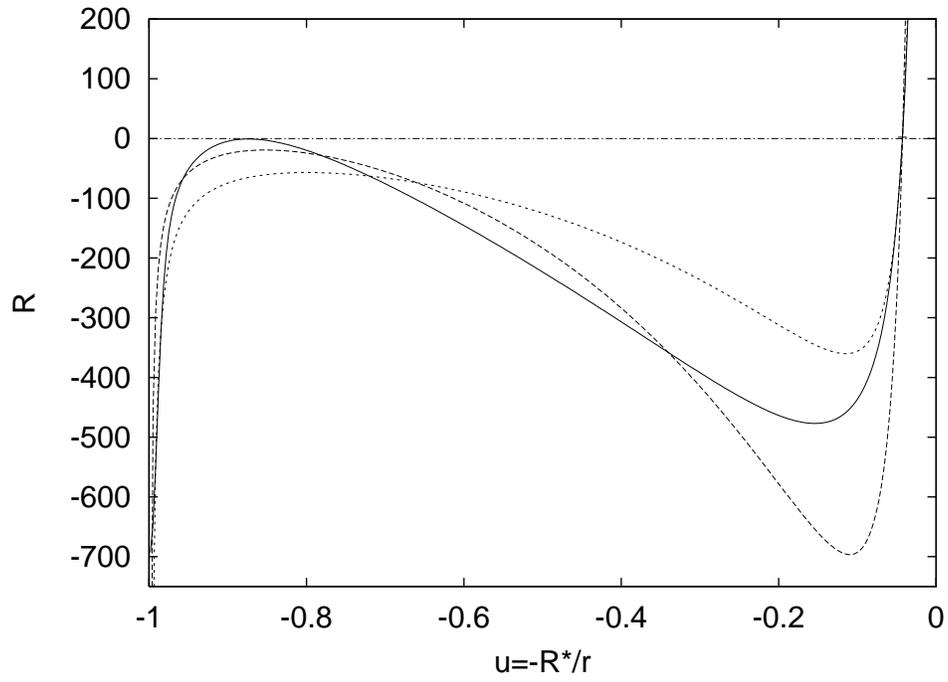}
\caption{Function $R(u,Z)$ from full numerical calculations, continuous line;
$R_{app}(u)$ for $\beta=1$ long--dashed line, and $R_{app}(u)$ for $\beta=2$
short--dashed line.  \label{fig_RR_be}} 
\end{figure}

In order to validate the $R_{app}(u)$ function, we have plotted in figure
\ref{fig_RR_be} the function $R(u,Z)$ from the results of the full numerical
calculations (continuous--line) and compared it with the $R_{app}(u)$ functions
for two different values of the $\beta$ parameter
(dashed--lines).\notetoeditor{Please place figure fig\_RR\_be.eps here} The
behavior of $R(u,Z)$ and $R_{app}(u)$ in the region far away in the wind ($u
\gtrsim -0.05$) is similar for both values of $\beta$, confirming that, also
for this new solution, the last root of the approximate function $R_{app}(u)$ gives an 
accurate value for the location of the critical point, $r_{crit}$, and through eq. (\ref{2.10}) and eq.(\ref{2.11}) the eigenvalue, $C'(\dot{M})$, for the fast--rotational case.

\subsection{$v_{\infty}$ and $\dot{M}$ from slow solutions. \label{sub-VM}}
As de Araujo (1994) and Stee at al. (1995, and references therein) argued,
it is possible to obtain equatorial solutions with lower terminal
velocities when the wind is driven mainly by thin lines, e.g., using a lower value
of the line force parameter $\alpha $. Thus, we have investigated the value of
$v_{rot}$ that makes that the standard m--CAK critical point disappears
remaining only the slow solution.
\notetoeditor{Please insert table \ref{tab5} here here}

Table (\ref{tab5}) summarises the value of this minimum rotational velocity
from the $R_{app}$ function for the existence of the slow solution. The values
of $\alpha $ and $k$  are taken from de Araujo(1994) and $\delta=0$ has been
used. The values of $v_{\infty}$ and $\dot{M}$ are coming from full numerical
solutions. This result suggests that in order to obtain a better modeling of
this slow--solution a self--consistent calculation (Pauldrach 1989) of the line
force parameters is necessary. This table (\ref{tab5}) shows, that the
thinner the lines drive the wind (lower $\alpha$), the lower is the required
rotational velocity for slow solution and therefore the lower is the terminal
velocity and mass loss rate. This table also shows a weak dependence of the
$\beta$--parameter on the minimal rotational velocity.

\section{Disk Formation in Be Stars. \label{sec-DF}}
\begin{figure}
\plotone{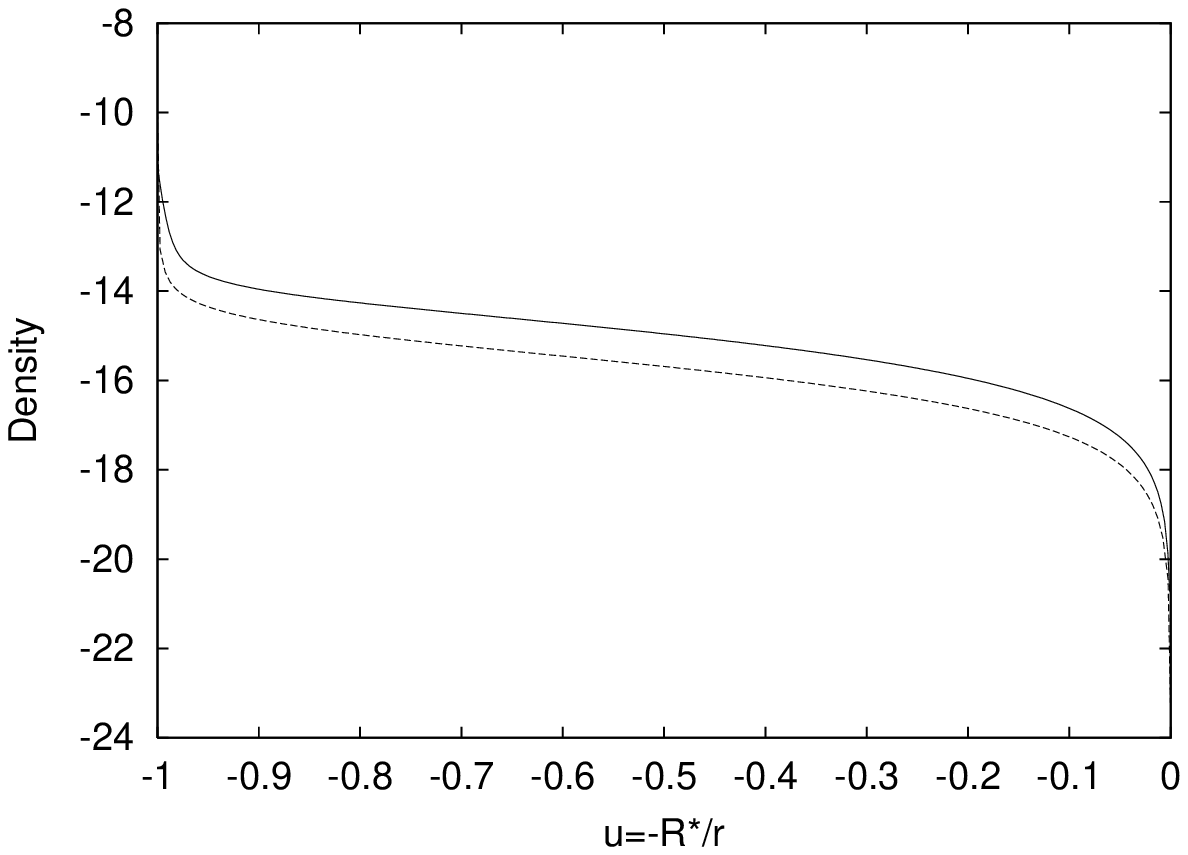}
\caption{Wind Density in $g\;cm^{-3}$ versus $u$ from full numerical
calculations  for the equatorial direction (continuous line) and polar direction
(dashed--line). \label{fig_dens_be}} 
\end{figure}

For our model is natural to think that a fast--rotating star can have both wind
solutions, e.g., Be Stars. 
The polar direction is equivalent to the non--rotational case, while the
equatorial flow corresponds to the fast--rotational case. In figure 
\ref{fig_dens_be} the density profile for both directions, i.e., polar
(dashed--line) and equatorial (continuous--line) are plotted against the
inverse radial coordinate $u$.

\notetoeditor{Please place figure fig\_dens\_be.eps here}
The ratio between the equatorial and polar densities are about a factor $100$
very close to the star's surface but mainly a factor $10$ for the whole wind.
This density quotient value is still a factor 10 smaller than the observational values. The disk
aperture angle occurs when the m--CAK critical point disappears from the
$R(u,Z)$ function, so this angle will depend strongly on the assumed value of
the line--force parameter $\alpha$. A detailed '2-D' (latitude dependent) velocity and density
profile and line formation is the subject of a forthcoming article.

\section{Conclusions \label{sec-CN}} 
After a suitable change of coordinates, we show that besides the 
standard m--CAK critical point there exists a second one when the star's 
rotational speed is taken into account for a hot star with radiation driven wind. 
Using a simple $\beta$--field approximation we developed an analytical
description for the location of the critical point and the value of the mass
loss rate of the star. Our solution remains within a 1\% confidence when compared with numerical calculations. This point is very important when trial solutions for numerical codes are selected. 

For the case of extreme high rotational speed (applied to a Be star), we have studied new solutions. Here exists only one critical point and numerical wind solutions were obtained. This {\it new} solution gives very slow terminal velocities (in the range $\sim 190$ $km sec^{-1}$ to $\sim 450$ $km sec^{-1}$, depending on the assumed values of the line force parameters) and the density of the wind is about 10 times denser than the standard m--CAK wind solution. These results indicate that new wind solution of the standard m--CAK model may be important in forming the disk in Be Stars. But to improve this model for Be stars, it is necessary that the following effects be to taken into account:  the dependence of the radius and temperature as a function of rotational speed and latitude, a self-consistent treatment of the line force parameters, a detailed study of the influence of the boundary conditions, a re-analysis of angular momentum conservation (including viscous forces) and the (ignored here) influence of magnetic fields. 

Currently a detailed study of the topology of the singular points is underway.
There are probably more solutions which involve more than one singular point as
the case of magnetic fields. 
 
\acknowledgments 
I would like to express my gratitude to N. Zamorano for his continuous support and critical comments on the manuscript.

This work has been partially supported by Universidad de Valpara\'{\i}so, 
internal project DIUV 14/00 and 15/03. 
 
\appendix 
\section{ Coordinate Transformation. \label{appA}} 
Here the basic steps toward equations (\ref{2.9a}), (\ref{2.9b}) and (\ref 
{2.9c}) are outlined. The reader should keep in mind the original derivation by
CAK.
 
The partial derivatives of $F(u,w,w\prime)$ (eq. \ref{2.9a}), with respect 
to $u,w$ and $w\prime$ are:  
\begin{equation} 
\frac{\partial F}{\partial u}= -\frac{2}{u^{2}} + a^{2}_{rot} - C^{\prime } 
\left( \frac{\partial CF}{\partial u}g + CF\frac{\partial g}{\partial u} 
\right) \;w^{-\delta }\;(ww\prime)^{\alpha }\;  \label{a1} 
\end{equation} 
\begin{equation} 
\frac{\partial F}{\partial w}=\left( 1+\frac{1}{w^{2}}\right) w^{\prime }- 
C^{\prime }\left( \frac{\partial CF}{\partial w}+\alpha \frac{CF}{w} 
-\delta \frac{CF}{w}\right) \;g\;w^{-\delta }\;(ww\prime)^{\alpha }  
\label{a2} 
\end{equation} 
\begin{equation} 
\frac{\partial F}{\partial w\prime}=\left( 1-\frac{1}{w^{2}}\right) w 
-C^{\prime }\left( \frac{\partial CF}{\partial w\prime}+\alpha 
\frac{CF}{w\prime}\right) \;g\;w^{-\delta }\;(ww\prime 
)^{\alpha }   
\label{a3} 
\end{equation} 
 
After using the new coordinate $Y=w\;w^{\prime }$ and $Z=w/w^{\prime }$, 
some derivative relation of the correction factor (see appendix \ref{appB}) and 
defining:  
\begin{equation} 
\frac{dg(u)}{du}=g(u) \; h(u)  \label{a4} 
\end{equation} 
where  
\begin{equation} 
h(u)=\delta \left( \frac{2}{u}-\frac{u}{\sqrt{1-u^{2}}\left(
1-\sqrt{1-u^{2}}\right) }\right)  \label{a5} 
\end{equation} 
the singularity condition ($w^{\prime }\;\partial F/\partial w^{\prime }=0$) 
reads:  
\begin{equation} 
\left(1-\frac{1}{YZ}\right)\;Y\;-C^{\prime} 
\;f_{2}(u,Z)\;g\;Z^{-\delta /2}\;Y^{\alpha -\delta /2}=0  \label{a6} 
\end{equation} 
here$\;f_{1}(u,Z)\equiv CF(u,Z)$,  
\begin{equation} 
f_{2}(u,Z)=\alpha \;f_{1}(u,Z) - u\;Z\;e(u,Z)  \label{a8} 
\end{equation} 
and $e(u,Z)$ is a function defined in appendix \ref{appB}. 
 
The regularity condition ($Z\;dF/du=0$) transform to:  
\begin{equation} 
\left(1+\frac{1}{YZ}\right)\;Y-C^{\prime} 
\;f_{3}(u,Z)\;g(u)\;Z^{-\delta /2}\;Y^{\alpha -\delta
/2}=+\frac{2Z}{u^{2}}-a_{rot}^{2}Z\label{a9} 
\end{equation} 
with  
\begin{equation} 
f_{3}(u,Z)=(3u+Z)\;Z\;e(u,Z)+f_{1}(u,Z)\;(h(u)\;Z+\alpha
-\delta )  \label{a10} 
\end{equation} 
 
\section{The Correction Factor\label{appB}} 
The Correction Factor is defined by: 
\begin{equation} 
CF=\frac{2}{1-\mu _{\ast}^{2}}\int_{\mu _{\ast}}^{1} \left( \frac{ 
(1-\mu ^{2})v/r+\mu ^{2}v^{\prime }}{v^{\prime }}\right) ^{\alpha }\mu d\mu . 
\label{eqa1} 
\end{equation}where $\mu _{\ast}^2={1-(R_{\ast}/r)^{2}}\;$and$\;$ $v^{\prime
}=dv/dr$.  Integrating \ref{eqa1} and changing the variables from $r,v\;$
to$\;$ $u, w$ $\;$($u=-R_{\ast }/r$, $w=v/a$, where $a\;$is the
thermal velocity), the finite disc correction factor transform to: 
\begin{equation} 
CF(u,w/w')=\frac{1}{1-\alpha
}\frac{1}{u^{2}}\frac{1}{(1+\frac{1}{u}\frac{w}{w^{\prime }})}\left[
1-\left( 1-u^{2}-u\frac{w}{w^{\prime}}\right)^{(1+\alpha )}\right]. 
\label{eqa2} 
\end{equation} where $w^{\prime }=dw/dr$. Due to the fact, that $CF$
depends on $u$ and {\it only} the quotient $Z\equiv w/w^{\prime }$, so defining
$\lambda$ as:
\begin{equation}
\lambda \equiv u \; (u+Z)
\label{eqa2b} 
\end{equation} we can re--write $CF(u,Z=w/w')$ as: 
\begin{equation} 
CF(\lambda)=\frac{1}{(1-\alpha)}\frac{1}{\lambda} \left[1-\left(1-\lambda
\right)^{(1+\alpha)} \right].  \label{eqa3} 
\end{equation} 
 
The partial derivatives of $CF$ with respect to $u,w,w^{\prime }$ are 
related to $\partial CF/\partial \lambda $ via the chain rule, namely:
\begin{equation} 
\frac{\partial CF}{\partial u}=\frac{\partial CF}{\partial \lambda }\frac{ 
\partial \lambda }{\partial u}.  \label{eqa4} 
\end{equation} 
\begin{equation} 
\frac{\partial CF}{\partial w}=\frac{\partial CF}{\partial \lambda }\frac{ 
\partial \lambda }{\partial w}.  \label{eqa5} 
\end{equation} 
\begin{equation} 
\frac{\partial CF}{\partial w^{\prime }}=\frac{\partial CF}{\partial \lambda  
}\frac{\partial \lambda }{\partial w^{\prime }}.  \label{eqa6} 
\end{equation} 
 
Defining $\;$ $e(\lambda )\equiv e(u,\lambda )=\partial CF/\partial \lambda $ 
\begin{equation} 
e(\lambda )=\frac{ \left( 1- \lambda \right)^{\alpha} - CF(\lambda)}{\lambda}. 
\label{eqa7} 
\end{equation} 
then (\ref{eqa4}),(\ref{eqa5}) and (\ref{eqa6}) are related to (\ref{eqa7}) 
by:  
\begin{equation} 
e(\lambda )=\frac{1}{2u+Z}\frac{\partial CF}{\partial
u}=\frac{w'}{u} \frac{\partial CF}{\partial w}=-\frac{w^{\prime
}}{uZ}\frac{\partial CF}{ 
\partial w^{\prime }}.  \label{eqa8} 
\end{equation} 
Approximating $Z$ by a $\beta$--field ($Z=(1+u)/\beta$), 
we obtain $CF$ and $e$ {\it only }as functions of $u$.

%
%
\clearpage 

\begin{deluxetable}{crrrrcrrrr} 
\tablecolumns{10} 
\tablewidth{0pc} 
\tablecaption{The location of the critical point, $r_{crit}$, and the 
mass-loss rate, $\dot{M}$, calculated by the analytical description for the 
CAK-model and compared against full numerical calculations. \label{tab-CAK}}
\tablehead{ 
\colhead{${v_{rot}}/{v_{breakup}}$}  
& \multicolumn{4}{c}{$r_{crit}/R_{\ast}$} 
& \colhead{}   
& \multicolumn{4}{c}{$\dot{M}$  $(10^{-6} M_{\sun}$ $year^{-1})$}\\
\cline{2-5} \cline{7-10} 
\colhead{}
& \colhead{$\beta=1/2$}  & \colhead{$\beta=1$}  & \colhead{$\beta=2$} 
& \colhead{Numeric} 
& \colhead{}   
& \colhead{$\beta=1/2$}  & \colhead{$\beta=1$}  & \colhead{$\beta=2$} 
& \colhead{Numeric} 
}
\startdata 
0.0 &  1.500 &  2.000 &  3.000 &  1.562 & & 3.074 & 3.086 & 3.093 & 3.083  \\ 
 
0.3 &  6.125 &  6.211 &  6.537 &  6.186 & & 3.127 & 3.128 & 3.131 & 3.128  \\ 

0.5 & 10.175 & 10.220 & 10.398 & 10.100 & & 3.163 & 3.165 & 3.167 & 3.164  \\ 

0.7 & 14.233 & 14.264 & 14.385 & 14.507 & & 3.201 & 3.202 & 3.205 & 3.203  \\ 

0.9 & 18.295 & 18.318 & 18.409 & 18.555 & & 3.240 & 3.241 & 3.243 & 3.243  \\
 
\enddata 
\end{deluxetable} 
\clearpage 

\begin{deluxetable}{crrrrcrrrr} 
\tablecolumns{10} 
\tablewidth{0pc} 
\tablecaption{Analytical calculation for $r_{crit}$ and $\dot{M}$ for the
m-CAK model compared against full numerical calculations. \label{tab-OB}}
\tablehead{ 
\colhead{${v_{rot}}/{v_{breakup}}$}  
& \multicolumn{4}{c}{$r_{crit}/R_{\ast}$} 
& \colhead{}   
& \multicolumn{4}{c}{$\dot{M}$  $(10^{-6} M_{\sun}$ $year^{-1})$}\\
\cline{2-5} \cline{7-10} 
\colhead{}
& \colhead{$\beta=0.8$}  & \colhead{$\beta=1$}  & \colhead{$\beta=2$} 
& \colhead{Numeric} 
& \colhead{}   
& \colhead{$\beta=0.8$}  & \colhead{$\beta=1$}  & \colhead{$\beta=2$} 
& \colhead{Numeric} 
}
\startdata 
0.0 & 1.026 & 1.034 & 1.084 & 1.033 & & 2.089 & 2.131 & 2.364 & 2.129  \\
 
0.3 & 1.027 & 1.035 & 1.088 & 1.036 & & 2.235 & 2.280 & 2.520 & 2.280  \\ 

0.5 & 1.029 & 1.038 & 1.098 & 1.040 & & 2.566 & 2.618 & 2.903 & 2.623  \\ 

0.7 & 1.036 & 1.048 & 1.129 & 1.051 & & 3.372 & 3.437 & 3.782 & 3.452  \\ 
\enddata 
\end{deluxetable} 
\clearpage 

\begin{deluxetable}{crrrrcrrrr} 
\tablecolumns{10} 
\tablewidth{0pc} 
\tablecaption{
Analytical calculation for $r_{crit}$ and $\dot{M}$ for a CSPN compared against
full numerical calculations. \label{tab-CSPN}} 
\tablehead{ 
\colhead{${v_{rot}}/{v_{breakup}}$}  
& \multicolumn{4}{c}{$r_{crit}/R_{\ast}$} 
& \colhead{}   
& \multicolumn{4}{c}{$\dot{M}$  $(10^{-9} M_{\sun}$ $year^{-1})$}\\
\cline{2-5} \cline{7-10} 
\colhead{}
& \colhead{$\beta=1$}  & \colhead{$\beta=1.5$}  & \colhead{$\beta=2$} 
& \colhead{Numeric} 
& \colhead{}   
& \colhead{$\beta=1$}  & \colhead{$\beta=1.5$}  & \colhead{$\beta=2$} 
& \colhead{Numeric} 
}
\startdata 
0.0 & 1.042 & 1.069 & 1.102 & 1.074 & & 4.638 & 4.881 & 5.140 & 4.916  \\
 
0.3 & 1.044 & 1.073 & 1.108 & 1.080 & & 4.882 & 5.141 & 5.418 & 5.189  \\ 

0.5 & 1.050 & 1.082 & 1.123 & 1.092 & & 5.424 & 5.716 & 6.024 & 5.791  \\ 

0.7 & 1.064 & 1.107 & 1.163 & 1.126 & & 6.678 & 7.021 & 7.372 & 7.137  \\ 
\enddata 
\end{deluxetable}  
\clearpage 

\begin{deluxetable}{ccccccccc} 
\tablecolumns{9} 
\tablewidth{0pc} 
\tablecaption{
The value of the minimum rotational rate for the 
case when only one solution exists. \label{tab5}} 
\tablehead{ 
\colhead{$\alpha$}  
& \colhead{$k$} 
& \colhead{}
& \multicolumn{3}{c}{$v_{rot}/v_{breakup}$} 
& \colhead{}
& \colhead{$v_{\infty}$}
& \colhead{$M_{\odot}$}\\
\cline{4-6}  
\colhead{}   
& \colhead{}
& \colhead{}   
& \colhead{$\beta=1$}  & \colhead{$\beta=1.5$}  & \colhead{$\beta=2$} 
& \colhead{}
& \colhead{($km$ $sec^{-1}$)} 
& \colhead{($10^{-9}M_{\sun}$ $year^{-1}$)}
}
\startdata 
0.5 & 0.3 & &0.88 & 0.87 & 0.86 & &454 & 6.86  \\

0.4 & 0.7 & &0.82 & 0.81 & 0.79 & &375 & 3.72  \\  

0.3 & 1.5 & &0.75 & 0.72 & 0.69 & &307 & 1.15  \\  

0.2 & 3.2 & &0.64 & 0.56 & 0.56 & &245 & 0.13  \\  

0.1 & 6.0 & &0.44 & 0.38 & 0.34 & &187 & .0001 \\
\enddata 
\end{deluxetable}  
\clearpage 
%
%

\clearpage 
 
\begin{references} 
\reference{AB82}  Abbott, D.C., 1982, \apj, 259, 282
\reference{BoMA}  Boyd, C.J. and Marlborough, J.M., 1991 , \apj, 369, 191 
\reference{BJ95}  Bjorkman, J.E., 1995, \apj, 453, 369
\reference{BC93}  Bjorkman, J.E. and Cassinelli, J.P., 1993, \apj, 409, 429 
\reference{CA74}  Castor, J.I., 1974, \apj, 183, 273
\reference{CA79}  Castor, J.I., 1979, Mass Loss and Evolution of O-Type Stars,
IAU Symp 83, eds. Conti,P.S., and de Loore, C.W., Dordrecht: Reidel, 175 
\reference{CAK}   Castor, J.I., Abbott,D.C. and Klein,R. 1975, \apj, 195, 157 
\reference{CS98}  Cassinelli, J., 1998, B[e] stars : Proceedings of the Paris
workshop held 9-12 June, 1997. Edited by A. M. Hubert and C.Jaschek. Dordrecht;
Boston: Kluwer Academic Publishers, (Astrophysics and Space Science library; v. 233), p.177
\reference{DA95}  de Araujo, F.X., 1995 , \aap, 298, 179 
\reference{AF89}  de Araujo, F.X. and de Freitas Pacheco, J.A., 1989, \mnras, 267, 501 
\reference{AFP94} de Araujo, F.X., de Freitas Pacheco, J.A. and Petrini, N., 1994 ,\mnras, 267, 501 
\reference{FA86}  Friend, D. and Abbott,D.C., 1986, \apj, 311, 701 
\reference{FM84}  Friend, D. and MacGregor,K.B., 1984, \apj, 282, 591
\reference{HA94}  Hanuschik, R. W., 1994, in Pulsation, Rotation and Mass Loss in Early-type Stars,IAU symposium 162, Eds L.A. Balona, H.F. Henrichs and J.M.Contel (Kluwer Academic Publishers, Dordrecht), p.265
\reference{KO92}  Koninx, J.P.M. and Hearn, A.G., 1992, \aap, 263, 208 
\reference{KPPA}  Kudritzki,R.P., Pauldrach, A., Puls,J. and Abbott,D.C., 
1989, \aap, 219, 205 
\reference{KP00}  Kudritzki,R.P. and Puls,J., 2000, \araa 38, 613 
\reference{KU97}  Kudritzki,R.P., Mendez,R.H., Puls,J. and McCarthy,J.K., 1997, Planetary nebulae, IAU Symp. 180, eds. H.J. Habing and H.J.G.L.M. Lamers. Dordrecht: Kluwer Academic Publishers.
\reference{NT88}  Nobili, L. and Turolla, R., 1988, \apj 333, 248
\reference{MZ84}  Marlborough, J.M. and Zamir, M., 1984, \apj, 276, 706 
\reference{ow94}  Owocki, S.P., Cranmer, S.R. and Blondin, J.M., 1994, \apj, 424, 887 
\reference{ow96}  Owocki, S.P., Cranmer, S.R. and Gayley, K.G., 1996, \apj, 472, 115 
\reference{PA89}  Pauldrach, A., 1989, \aap, 183, 295 
\reference{PA88}  Pauldrach, A., Puls, J., Kudritzki, R.P., Mendez, R.H. and Heap, S. R. 1988, \aap, 207, 123 
\reference{PPK}   Pauldrach, A., Puls,J. and Kudritzki,R.P., 1986, \aap, 164, 86 
\reference{pf86}  Poe, C., and Friend, D., 1986, \apj, 311, 317 
\reference{PM78}  Poeckert, R. and Marlborough, J.M., 1978, \apj, 220, 940 
\reference{WM94}  Waters, L.B.F.M. and  Marlborough, J.M., 1994, in Pulsation, Rotation and Mass Loss in Early-type Stars,IAU symposium 162, Eds L.A. Balona, H.F. Henrichs and J.M.Contel (Kluwer Academic Publishers, Dordrecht), p.399
\end{references}
\end{document}